\documentclass[10pt, conference, letterpaper]{IEEEtran}
\pdfoutput=1
\usepackage[utf8]{inputenc}
\usepackage{csquotes}
\usepackage[hyphens]{url}
\usepackage{xspace}
\usepackage{expl3}
\usepackage{todonotes}
\usepackage{amsmath}
\usepackage{amssymb}
\usetikzlibrary{calc}
\usepackage{adjustbox}
\usepackage{eurosym}
\usepackage{siunitx}
\usepackage{pifont}
\usepackage{amsthm}
\usepackage{tikz}
\usetikzlibrary{shapes}
\usetikzlibrary{positioning}
\usepackage{graphicx}
\usepackage{subfig}
\usepackage{hyperref}
\hypersetup{hidelinks,
  colorlinks=true,
  allcolors=black,
  pdfstartview=Fit,
  breaklinks=true}
\usepackage[hypcap=true]{caption}

\usepackage[%
 backend=biber,
 style=ieee,
 url=false,
 isbn=false,
 doi=false,
 maxcitenames=1,
 mincitenames=1,
 maxbibnames=5,
]{biblatex}

\AtEveryBibitem{\clearlist{location}}
\AtEveryBibitem{\clearfield{pages}}
\AtEveryBibitem{\clearfield{series}}
\AtEveryBibitem{\clearfield{month}}

\newcommand*{\eg}{e.g.\@\xspace}
\newcommand*{\ie}{i.e.\@\xspace}
\newtheorem{theorem}{Theorem}

\begin{document}
\title{Monetary Stabilization in Cryptocurrencies---Design Approaches and Open Questions}

\author{
\IEEEauthorblockN{%
Ingolf G.A. Pernice\IEEEauthorrefmark{1}\IEEEauthorrefmark{2} \quad\qquad%
Sebastian Henningsen\IEEEauthorrefmark{1}\IEEEauthorrefmark{2} \quad\qquad%
Roman Proskalovich\IEEEauthorrefmark{1}\IEEEauthorrefmark{2}\IEEEauthorrefmark{3}\\%
Martin Florian\IEEEauthorrefmark{1}\IEEEauthorrefmark{2} \quad\qquad%
Hermann Elendner \quad\qquad%
Björn Scheuermann\IEEEauthorrefmark{1}\IEEEauthorrefmark{2}%
}
\IEEEauthorblockA{~\\
\IEEEauthorrefmark{1}Weizenbaum-Institute for the Networked Society\\
\IEEEauthorrefmark{2}Humboldt-Universität zu Berlin\\
\IEEEauthorrefmark{3}Belarusian State University\\
}}
\maketitle
\begin{abstract}
The price volatility of cryptocurrencies is often cited as a major hindrance to their wide-scale adoption.
Consequently, during the last two years, multiple so called \emph{stablecoins} have surfaced---cryptocurrencies focused on maintaining stable exchange rates.
In this paper, we systematically explore and analyze the stablecoin landscape.
Based on a survey of 24 specific stablecoin projects, we go beyond individual coins for extracting general concepts and approaches.
We combine our findings with learnings from classical monetary policy, resulting in a comprehensive taxonomy of cryptocurrency stabilization.
We use our taxonomy to highlight the current state of development from different perspectives and show blank spots.
For instance, while over 91\% of projects promote 1-to-1 stabilization targets to external assets, monetary policy literature suggests that the smoothing of short term volatility is often a more sustainable alternative.
Our taxonomy bridges computer science and economics, fostering the transfer of expertise.
For example, we find that 38\% of the reviewed projects use a combination of exchange rate targeting and specific stabilization techniques that can render them vulnerable to speculative economic attacks---an avoidable design flaw.
\end{abstract}

\section{Introduction}
\label{sec:intro}

Although cryptocurrencies like Bitcoin have gained a lot of attention over the last years, they have not been adopted as standard means of payment.
The large fluctuations in coin prices are often cited as one of the main reasons for everyday users' reluctance~\cite{saito_how_2018,yermack2015bitcoin,basecoin}.
An increasing number of cryptocurrencies consequently devote themselves to maintaining a stable price.
These so-called ``stablecoins'' promise the best of both worlds: a (permissionless) cryptocurrency such as Bitcoin combined with the price stability of traditional fiat currencies such as the US Dollar.

What is the current state of this development?
What has been done, and what can be done to ensure stability?
In this paper, we systematically explore and analyze the stablecoin landscape.
We go beyond individual projects and instead provide an abstract overview of concepts merged with approaches from traditional monetary policy.

We surveyed white papers, websites and, when available, price data of 24 stablecoin projects.
While the short-lived nature of most cryptocurrencies quickly makes any survey of existing coins obsolete, the abstract perspective allows us to reason about fundamental properties, risks and limitations of stability techniques in practice.
We used the union between generalized design features and monetary theory for developing a comprehensive taxonomy on stabilization approaches for cryptocurrencies.
Our taxonomy tackles three broad questions that are reflected in the structure of this paper:
\begin{enumerate}
  \item Which types of practical techniques are used to achieve stability? (Sec.~\ref{sec:stab_techniques})
  \item In what way can the value of a cryptocurrency be linked to that of an external currency (\eg, the US Dollar (USD), the Euro (EUR))? (Sec.~\ref{sec:era})
  \item What is the stabilization target (\eg, exchange rate to USD, inflation, etc.)? (Sec.~\ref{sec:mr})
\end{enumerate}
We use our taxonomy to highlight the current state of development from different dimensions and show blank spots.
As our taxonomy bridges computer science and economics it allows for transfer of expertise.
This not only leads to the detection of risks but also reveals avenues for future research.

On a more detailed level, we find that almost~\SI{38}{\percent} of surveyed coins promote a problematic combination of exchange rate targeting and techniques for reducing the coin supply (using either limited reserves or a potentially unlimited supply of self-issued tokens).
While more research is encouraged, there is strong indication that this might render them vulnerable to speculative attacks, \ie, scenarios in which investors deliberately apply market pressure to push the price of a coin below the stable value to make a profit.

Furthermore, existing economic literature suggests that soft pegs are not maintainable in the long run and that more sustainable arrangements such as \emph{smoothing} of short term variations or \emph{hard pegs} are preferable.
When it comes to the state of developments, simple tokenization of national currency is the most popular technique.
More sophisticated techniques have been planned for implementation, however, many face inherent challenges such as not allowing a permanent reduction of the money supply.

Although we heavily focus on the economic perspective, we also point out technical challenges. For example, almost all surveyed stablecoins rely on a trusted price feed and therefore a functioning decentralized oracle.
This assumption is problematic, as existing research~\cite{DBLP:journals/corr/abs-1808-00528,DBLP:journals/corr/PetersonK15,DBLP:conf/ccs/ZhangCCJS16} does not solve the decentralized oracle problem for arbitrary values and a general solution might be impossible due to the lack of strong identities~\cite{douceur2002sybil} or missing incentive-compatibility.
While touching on hard technical considerations only lightly, we argue that the proposed viewpoint provides a valuable transfer of knowledge between economics and computer science by opening up new perspectives on a predominantly technical discussion.

\section{Related work}
\label{sec:rw}

Monetary stability in cryptocurrencies has barely been studied by the scientific community.
Iwamura et al.~\cite{iwamura_can_2014,saito_how_2018,saito_how_2018} propose a combination of dynamic mining reward and automatic inflation of coins.
In a different approach, Caginalp et al. argue~\cite{caginalp2018opinion} that since cryptocurrencies have no underlying value measured by ``traditional techniques'' that are used to value stocks, bonds or derivatives, new models are necessary.
Following this idea, Caginalp~\cite{caginalp_dynamical_2018} uses asset flow equations to model the price of cryptocurrencies and derive conditions under which the models differential equations stabilize.
In contrast to proposing in-depth designs of novel stabilization approaches, we focus on surveying existing projects and outlining principal features of the design space.

Another branch of scientific research concerns itself with \emph{central bank digital currency (CBDC)}~\cite{danezis_centrally_2015,bech_central_2017,koning2016fedcoin,bjerg_designing_2017}.
We deliberately chose not to cover this topic, since the central bank, as the central actor, remains in control of both monetary policy and mining.
Effectively, this creates another form of national money, leaving monetary policy aspects mostly unchanged.

There already exists variety of (non-scientific) classifications and stablecoin lists,\footnote{  For example:\\
\url{https://github.com/sdtsui/awesome-stablecoins}\\
\url{https://stablecoinindex.com/}\\
\url{https://media.consensys.net/the-state-of-stablecoins-2018-79ccb9988e63}\\
\url{https://cryptoinsider.21mil.com/stablecoins-everything-need-know/}\\
\url{https://hackernoon.com/stablecoins-designing-a-price-stable-cryptocurrency-6bf24e2689e5}%
}
as well as prominent criticism of the concept itself.\footnote{ \url{https://prestonbyrne.com/2018/03/22/stablecoins-are-doomed-to-fail/}}
Although valuable and highly informative, in our analysis we take a broader, more structured and systematic approach, by stepping away from specific projects and towards the underlying concepts.

\section{Survey Methodology}
\label{sec:intro_frame}

Exploring and understanding the stablecoin landscape is a tedious task but a prerequisite for any further insights.
Due to the short lived nature of many coins, any such perspective is necessarily a momentary snapshot of current projects.
Therefore, we avoid reasoning about individual coins and only use them as examples for abstract approaches instead.\footnote{
  Classification details for individual projects can be referred to in Appendix~\ref{appendix:surveyed_projects}.
}
In~\cite{kim2018crypto}, stablecoins are identified as cryptocurrencies ``whose values are pegged to [\ie stabilized relative to] some other fiat money or asset with inherent value''.
This definition of stablecoins, however, is exceedingly narrow.
In traditional economics, stability can go beyond a long-term link to an foreign currency or some asset.
As a consequence, we broaden the definition to cryptocurrencies ``with mechanisms to mitigate fluctations in their purchasing power".\footnote{
  \emph{Purchasing power of a currency} describes how many units of certain goods, services or other currencies it can buy.
}

The scope of our analysis is limited to stablecoins that are (i) permissionless, (ii) intended for general use as a currency, and (iii) provide a whitepaper and website.
By permissionless~\cite{DBLP:conf/cvcbt/WustG18} we mean any coin or token that runs on a permissionless Blockchain---specifically including IOU-Tokens such as Tether~\cite{tether}.
We exclude central bank digital currencies, pure utility tokens and stablecoins without a website or corresponding white paper.
At the time of writing we identified 24 projects that fit these criteria.\footnote{
  Some projects issue stablecoins pegged to different national currencies. In these cases we exclusively address the coin pegged to the USD.
}
Of these projects, 13 are launched and traded on exchanges.
As an overview of their performance, Table~\ref{tbl:desc} summarizes the mean, standard deviation, minimum and maximum price (in USD) of each launched coin according to data gathered from \url{https://coinmarketcap.com}.
It can be observed that the projects show divergent performance and general statistic characteristics.
This is due to the fact that each coin chooses a different strategy to stabilize its value.

We analyze approaches to achieve price stability using three classifications based on the monetary regimes proposed in~\cite{mishkin1999international},
an International Monetary Fund (IMF) study on exchange-rate arrangements~\cite{international2016exchange}
and practice of stabilization techniques by major central banks~\cite{fed,ecb}.

In the following section, we investigate practical stabilization techniques proposed.

\begin{center}
  \begin{table}
  	\caption{Basic descriptive statistics for available daily USD-prices of stablecoin projects until the 7th of February 2019.}
    \begin{tabular}{@{\extracolsep{-5pt}} cccccc} 
\\[-1.8ex]\hline 
\hline \\[-1.8ex] 
Projects & Obs. & Mean & Min. & Max. & Std. Dev. \\ 
\hline \\[-1.8ex] 
NuBits (Nubits) & $1587$ & $0.819$ & $0.030$ & $1.264$ & $0.332$ \\ 
BitShares (BitUSD) & $1525$ & $1.016$ & $0.680$ & $1.600$ & $0.076$ \\ 
Tether (USDT) & $1429$ & $1.000$ & $0.914$ & $1.058$ & $0.010$ \\ 
Karbo (Karbo) & $904$ & $0.299$ & $0.005$ & $2.066$ & $0.418$ \\ 
Minex Coin (Minex Coin) & $458$ & $10.991$ & $0.549$ & $56.586$ & $10.201$ \\ 
Maker (Dai) & $404$ & $1.002$ & $0.939$ & $1.053$ & $0.010$ \\ 
Trusttoken (TrueUSD) & $335$ & $1.006$ & $0.985$ & $1.132$ & $0.012$ \\ 
Digix (Digix Gold Token) & $263$ & $42.075$ & $36.243$ & $50.207$ & $2.294$ \\ 
Sythetix (SUSD) & $205$ & $0.989$ & $0.867$ & $1.029$ & $0.018$ \\ 
Stasis (EURS) & $188$ & $1.143$ & $1.086$ & $1.260$ & $0.025$ \\ 
Centre (USD Coin) & $118$ & $1.013$ & $0.983$ & $1.037$ & $0.008$ \\ 
Stronghold (USDS) & $46$ & $1.016$ & $0.947$ & $1.076$ & $0.021$ \\ 
USC (USC) & $32$ & $0.912$ & $0.656$ & $1.027$ & $0.152$ \\ 
\hline \\[-1.8ex] 
\end{tabular} 
    \label{tbl:desc}
  \end{table}
  \vspace{-4mm}
\end{center}

\section{Stabilization techniques}
\label{sec:stab_techniques}

Fundamentally, all stabilization techniques are based on the elementary economic model of supply and demand.
The \emph{price of a currency}\footnote{
The price of a currency describes how many units of other currencies are given in exchange for it.
The price can also be expressed in terms of other goods or services. 
Thus, in presented context the term can be seen as an equivalent to the currency's \emph{exchange rate} and \emph{purchasing power}.
}
can be modeled as the level at which its supply and demand meet each other on the market.
A change in price is therefore due to changes in supply and/or demand---to maintain stability, any such change has to be counteracted.

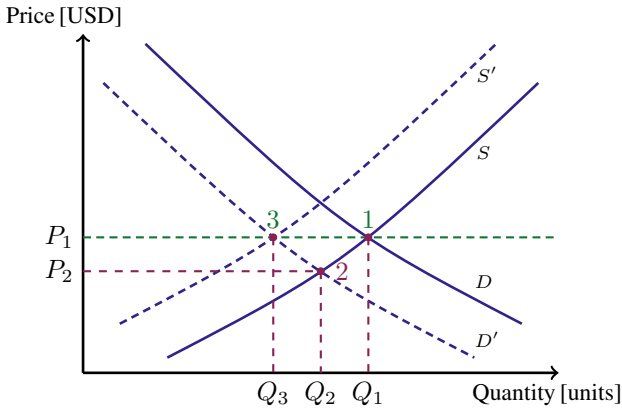
\begin{figure}[h]
\centering
\usetikzlibrary{calc, through}
\tikzset{
  every point/.style = {radius={\pgflinewidth}, opacity=1, draw, solid, fill=white},
  pt/.pic = {
    \begin{pgfonlayer}{foreground}
      \path[every point, #1] circle;
    \end{pgfonlayer}
  },
  point/.style={insert path={pic{pt={#1}}}}, point/.default={},
  point name/.style = {insert path={coordinate (#1)}}
}
\definecolor{CMQBlueA}{HTML}{332288}  
\definecolor{CMQBlueB}{HTML}{88CCEE}  
\definecolor{CMQBlueC}{HTML}{44AA99}  
\definecolor{CMQGreenA}{HTML}{117733} 
\definecolor{CMQGreenB}{HTML}{999933} 
\definecolor{CMQGreenC}{HTML}{DDCC77} 
\definecolor{CMQRedA}{HTML}{CC6677}   
\definecolor{CMQRedB}{HTML}{882255}   
\definecolor{CMQRedC}{HTML}{AA4499}   
\definecolor{CMQGray}{HTML}{DDDDDD}   

\pgfdeclarelayer{background}%
\pgfdeclarelayer{foreground}%
\pgfsetlayers{background,main,foreground}%

\begin{tikzpicture}[]
  \def \debugPoint{}%
  \def \colorCoor   {black!100!white}
  \def \lineCoorW   {1.0pt}
  \def \lineRedW    {0.8pt}
  \def \lineGreenW  {0.8pt}
  \def \lineCurvesW {1.0pt}
  \def \picW        {\linewidth*29.5/31}%
  \def \picH        {\paperheight*5.0/31}%
  \def \coorW       {\picW*0.75}
  \def \coorWH      {\coorWH*1/2}
  \def \coorH       {\picH}
  \def \coorHH      {\coorH*1/2}
  \def \distW       {\coorW*0.1}
  \def \distH       {\coorH*0.1}
  \coordinate (cen)      at ($(0,0)     + (-1cm            , 0.0             )$) (cen)    [\debugPoint];%
  \coordinate (CCLD)     at ($(cen)     + (-\coorW*0.5     ,-\coorH*0.5      )$) (CCLD)   [\debugPoint];%
  \coordinate (CCLU)     at ($(cen)     + (-\coorW*0.5     , \coorH*0.5      )$) (CCLU)   [\debugPoint];%
  \coordinate (CCRD)     at ($(cen)     + ( \coorW*0.5     ,-\coorH*0.5      )$) (CCRD)   [\debugPoint];%
  \coordinate (CCRU)     at ($(cen)     + ( \coorW*0.5     , \coorH*0.5      )$) (CCRU)   [\debugPoint];%
  \coordinate (CDB)      at ($(cen)     + ( 0.0            ,-\coorH*0.5      )$) (CDB)    [\debugPoint];%
  \coordinate (CDA)      at ($(CDB)     + (-\distW         , 0.0             )$) (CDA)    [\debugPoint];%
  \coordinate (CDC)      at ($(CDB)     + ( \distW         , 0.0             )$) (CDC)    [\debugPoint];%

  \coordinate (CLcen)    at ($(cen)     + ( 0.0            ,-\distH*2        )$) (CLcen)  [\debugPoint];%
  \coordinate (CLA)      at ($(CLcen)   + (-\coorW*0.5     , 0.0             )$) (CLA)    [\debugPoint];%
  \coordinate (CLB)      at ($(CLA)     + ( 0.0            , \distH          )$) (CLB)    [\debugPoint];%
  
  \coordinate (CRA)      at ($(cen)     + ( \coorW*0.5     ,-\distH*2        )$) (CRA)    [\debugPoint];%
  \coordinate (CRB)      at ($(CRA)     + ( 0.0            , \distH          )$) (CRB)    [\debugPoint];%
  
  \coordinate (CIA)      at ($(CLcen)   + (-\distW         , \distH          )$) (CIA)    [\debugPoint];%
  \coordinate (CIB)      at ($(cen)     + ( 0.0            ,-\distH*2        )$) (CIB)    [\debugPoint];%
  \coordinate (CIC)      at ($(CLcen)   + ( \distW         , \distH          )$) (CIC)    [\debugPoint];%
  
  \coordinate (CIOA)     at ($(CIB)     + (-\distW*2       , 0.0             )$) (CIOA)   [\debugPoint];%
  \coordinate (CIOB)     at ($(CIB)     + ( \distW*2       , 0.0             )$) (CIOB)   [\debugPoint];%
  
  \coordinate (CNLA)     at ($(cen)     + ( \coorW*0.35    ,-\coorH*0.405    )$) (CNLA)   [\debugPoint];%
  \coordinate (CNLB)     at ($(cen)     + ( \coorW*0.35    ,-\coorH*0.230    )$) (CNLB)   [\debugPoint];%
  \coordinate (CNLC)     at ($(cen)     + ( \coorW*0.35    , \coorH*0.150    )$) (CNLC)   [\debugPoint];%
  \coordinate (CNLD)     at ($(cen)     + ( \coorW*0.35    , \coorH*0.380    )$) (CNLD)   [\debugPoint];%

  \draw[\colorCoor, line width = \lineCoorW, -> ] (CCLD) -- (CCRD);
  \draw[\colorCoor, line width = \lineCoorW, -> ] (CCLD) -- (CCLU);
  \node[label={[above, xshift=-0.25cm, yshift=-0.15cm]{\small{}Price\,[USD]}}]      (NCLU) at (CCLU) {};
  \node[label={[below, xshift=-0.15cm, yshift=-0.15cm]{\small{}Quantity\,[units]}}] (NCRD) at (CCRD) {};
  
  \node[anchor = north] (NDB) at (CDB) {$Q_2$};
  \node[anchor = north] (NDA) at (CDA) {$Q_3$};
  \node[anchor = north] (NDC) at (CDC) {$Q_1$};
  
  \node[anchor = east ] (NLA) at (CLA) {$P_2$};
  \node[anchor = east ] (NLB) at (CLB) {$P_1$};
  
  \def \yShift {-2.5pt}
  \draw[CMQBlueA, line width = \lineCurvesW, shorten <=-2.3cm, shorten >=-3.1cm, solid, - ]          
    ($(CIB)+ ( 0.0cm , \yShift)$) to [bend right = 8pt] ($(CIC)+ ( 0.0cm , \yShift)$);
  \draw[CMQBlueA, line width = \lineCurvesW, shorten <=-2.3cm, shorten >=-3.2cm, densely dashed, - ] 
    ($(CIA)+ ( 0.0cm , \yShift)$) to [bend right = 8pt] ($(cen)+ ( 0.0cm , \yShift)$);
  \draw[CMQBlueA, line width = \lineCurvesW, shorten <=-3.1cm, shorten >=-2.3cm, densely dashed, - ] 
    ($(CIA)+ ( 0.0cm , \yShift)$) to [bend right = 8pt] ($(CIB)+ ( 0.0cm , \yShift)$);
  \draw[CMQBlueA, line width = \lineCurvesW, shorten <=-3.2cm, shorten >=-2.3cm, solid, - ]          
    ($(cen)+ ( 0.0cm , \yShift)$) to [bend right = 8pt] ($(CIC)+ ( 0.0cm , \yShift)$);

  \node[label={[align = flush right, yshift=-\coorH*0.075, font=\scriptsize]{$D'$}}] (NNLA) at (CNLA) {};
  \node[label={[align = flush right, yshift=-\coorH*0.075, font=\scriptsize]{$D\phantom{'}$}}]  (NNLB) at (CNLB) {};
  \node[label={[align = flush right, yshift=-\coorH*0.075, font=\scriptsize]{$S\phantom{'}$}}]  (NNLC) at (CNLC) {};
  \node[label={[align = flush right, yshift=-\coorH*0.075, font=\scriptsize]{$S'$}}] (NNLD) at (CNLD) {};
  
  \draw[CMQGreenA, line width = \lineGreenW, densely dashed, - ] (CLB) -- (CRB);
  
  \draw[CMQRedB, line width = \lineRedW, dashed, - ] (CLA) -- (CIB);
  \draw[CMQRedB, line width = \lineRedW, dashed, - ] (CIB) -- (CDB);
  \draw[CMQRedB, line width = \lineRedW, dashed, - ] (CIA) -- (CDA);
  \draw[CMQRedB, line width = \lineRedW, dashed, - ] (CIC) -- (CDC);
  
  \node[anchor = south] (NIA) at (CIA) {\color{CMQGreenA}$3$};
  \node[anchor = west, xshift=2pt ] (NIB) at (CIB) {\color{CMQRedB}$2$};
  \node[anchor = south] (NIC) at (CIC) {\color{CMQGreenA}$1$};
  \node[circle, minimum size = 3pt, text width = 3pt, inner sep = 0pt, outer sep = 0pt, fill=CMQRedB] (NIAB) at (CIA) {};
  \node[circle, minimum size = 3pt, text width = 3pt, inner sep = 0pt, outer sep = 0pt, fill=CMQRedB] (NIBB) at (CIB) {};
  \node[circle, minimum size = 3pt, text width = 3pt, inner sep = 0pt, outer sep = 0pt, fill=CMQRedB] (NICB) at (CIC) {};

\end{tikzpicture} 
  \caption{Examplary supply and demand model.}
  \label{fig:s_d_model}
\end{figure}

Fig.~\ref{fig:s_d_model} illustrates this concept, with price (in USD) on the y-axis and quantity (coins in this case) on the x-axis.
The solid $S$ and $D$ curves depict the money supply and demand, respectively.\footnote{
  Note that the specific shape of the curves is merely an example.
  It abstracts the market where a cryptocurrency is exchanged for goods, services or other currencies.
  For some cryptocurrency setups, in the short run, the money supply is independent of the price.
  This makes the money supply curve a vertical line that is shifted in the long run.
  The shape of the curve has no influence on the general rationale in the following explanations of stability techniques.
      }
Both curves intersect at $(Q_1, P_1)$, yielding an equilibrium quantity of $Q_1$ and price of $P_1$.
Stablecoins aim to maintain a constant price, say $P_1$ in this example.

Assume the demand decreases, \ie, users would purchase fewer coins at each price level, effectively shifting the demand curve to the left---from $D$ to $D'$.
The new equilibrium, the intersection of $D'$ and $S$ at $(Q_2, P_2)$, has a smaller quantity ($Q_2$) and lower price ($P_2$)---which violates the aim of maintaining a constant price $P_1$.
To recover, one can (i) increase demand (shift $D'$ to the right), (ii) decrease supply (shift $S$ to the left) or (iii) adjust both.
Especially for cryptocurrency systems, demand is much harder to influence directly and instantaneously, therefore, supply is often the target of choice.
This is depicted in Fig.~\ref{fig:s_d_model}, where supply is adjusted, shifting $S$ to $S'$ which yields an equilibrium of $(Q_3, P_1)$.
Here, the quantity of coins on the market ($Q_3$) is smaller, but their price in terms of US dollars is back to the desired level ($P_1$).

Whatever deviation from the initial price, the stable purchasing power can, theoretically, be restored by adjusting supply and demand.
Naturally, this model is a simplification and real-world examples are a lot more involved---however, it is helpful to analyze and classify techniques for maintaining stability.

In the following, we systematically investigate the techniques used by stablecoins to influence supply and demand and subsequently discuss potential risks and limitations.
The techniques are abstracted to underline their key features and to remove discrepancies in denominations used by different projects.
Furthermore, we compare these to techniques employed by traditional central banks.
In our analysis we identified six major techniques: (i) collateralization, (ii) interest rates, (iii) currency interventions, (iv) open market operations, (v) dynamic block reward and (vi) dynamically burned transaction fee.
The usage of those techniques is not mutually exclusive, a combination can be applied in practice.

\subsection{Tokenization of collateral}

Tokenization of collateral (or simply collateralization) links the coin supply to the demand, so that any change in the demand incentivizes market participants to change the supply accordingly.
Each stablecoin token is backed by a certain amount of \mbox{(crypto-)currencies}, assets or fiat money.
Users can create tokens by depositing an underlying backing, the so-called \emph{collateral} and can redeem (destroy) tokens to receive their collateral.
The entity which stores collateral might be a smart contract or centralized (as in the case of Tether~\cite{tether})---the limitations and drawbacks are discussed in Sec.~\ref{subsec:subsec:stab_techniques_focalpoints}.

The creation and destruction of coins through users provides a mechanism for supply adjustment.
On the one side, when demand increases, market participants can simply create new coins by depositing collateral, effectively increasing supply. 
Due to the excess demand, a coin might trade at a price higher than the value of the underlying collateral---in this case this arbitrage opportunity further incentivizes the creation of new coins.
On the other side, when demand decreases, supply can decrease as well by redeeming coins in exchange for their collateral and therefore destroying them.
Similarly, a coin might trade below the value of its collateral, creating arbitrage opportunities and therefore incentives to destroy coins.

Note that the described incentives (and therefore the success of this technique) rely on perfect transferability between coin and collateral: a coin can always be redeemed for its collateral and vice versa, without delays or any other friction.
A violation of this assumption in practice might make this technique less efficient and therefore a coin subject to price swings.

A number of assets have been proposed as collateral.
We distinguish three subcategory of collateralization: \emph{direct}, \emph{proxy} and \emph{self-collateralization}.

In \emph{direct collateralization}, each token is backed by the asset pegged to (\ie the asset it is stabilized against).
For example, if the goal is a stable exchange rate to the Euro, each token is backed by one Euro.
This design resembles the approach of fiat currencies such as the Bulgarian Lev backed by Euro or Djiboutian Franc backed by US dollars.
Examples of implemented projects include \emph{Stably}~\cite{stably} and \emph{Tether}~\cite{tether}.
The already implemented concepts show relatively stable exchange rates to the USD.
\emph{Stably}~\cite{stably} historically been able to maintain within a band of \SI{10}{\percent} around the peg, and \emph{Tether}~\cite{tether} even within a band of \SI{5}{\percent}.
There are, however, examples with larger deviations.

In \emph{proxy collateralization} each token is not backed by the targeted currency itself but instead by some other \mbox{(crypto-)currency}, asset or basket of assets.
Different from direct collateralization, there is a gap between collateral (\eg, Ether) and the stabilization target (\eg, USD):
falling prices of the collateral may lead to insufficient backing.

\emph{Self-collateralization} is a subform of proxy collateralization.
In this technique, another token which is issued within the ecosystem of the cryptocurrency itself is used as collateral.
The collateral risk is therefore elevated, since the fate of the ecosystem affects the stablecoin as well as its backing.
An already implemented example is the stablecoin \emph{BitUSD}~\cite{bitshares} which is backed by the token \emph{BitShares}~\cite{bitshares}.
While \emph{BitUSD}~\cite{bitshares} appeared relatively stable between 0.76 and 1.60 USD per \emph{BitUSD}~\cite{bitshares} for several years, it slumped below 0.67 USD in December 2018 when collateral prices declined.

For self and proxy collateralization, the gap between collateral and asset pegged to is mitigated with two (often combined) approaches:
first, requiring more collateral than necessary (\emph{over-collateralization}) and second, enforcing automatic re-collateralization (\emph{margin calls}).
In over-collateralization, more backing is required than the actual price goal of the token would suggest.
As an example, say a stable token backed with \emph{Bitcoin}\cite{bitcoin} should trade at 1 USD, then over-collateralization would require to deposit \emph{Bitcoin}\cite{bitcoin} worth 1.5 USD to create a token.
This allows for some volatility of the collateral without risking that tokens become undercollateralized, \ie, when the backing is worth less than the price goal of the token.

Margin calls are triggered, if the value of the collateral falls below a predetermined value, the ``margin'', in order to avoid undercollateralized tokens.
In a margin call either the creator of a token deposits more collateral or the collateral is offered for sale on the market in exchange for stable tokens.
Given sufficient liquidity on markets, this effectively rolls back the creation of a token and decreases its supply.

\subsection{Use of interest rates}

Interest rates are an instrument to guide a decentralized adjustment of the money supply.
For example, in the current real-world credit money system, most of the money is created when commercial banks issue loans to their clients~\cite{McLeay_et_al_2014}. 
The money stock decreases when loans are paid back or money in circulation is used to make deposits which lock money for a certain amount of time.
Central banks set and adjust the base interest rate to influence interest rates of the commercial banks.
The higher the rates, the smaller is the number of loans and the higher is the number of deposits in the system and the smaller the money supply becomes and vice versa for lower interest rates. 
The effectiveness of the technique ultimately depends on the decisions of the market participants to make \emph{deposits} and to take \emph{loans}.

\emph{Interest rates on deposits} are in some stablecoin projects denominated as \emph{parking} or \emph{locking} fees.
In this technique, users lock their coins in order to receive them back after a specific time with some additional reward (interest).
The interest is paid by the system, most often through the minting of new coins.
Higher interest rates make the currency more attractive for investors---demand increases.
At the same time, as a higher fraction of currency is locked in deposits, supply decreases; at least temporarily.
In the long run, supply only increases as deposits are paid back with interest rate.
Among others, ``Stableunit''~\cite{stableunit}, ``Minex Coin''~\cite{minex} and ``Nubits''~\cite{nubits} employ interest rates on deposits.

\emph{Interest rates on loans} are sometimes referred to as \emph{stability fees}.
Current implementations of loans in cryptocurrencies can be seen as a generalization of the collateralization technique: the stable token issued when depositing collateral is a loan on that collateral.
However, to get the collateral back a user has to return the stablecoin and may also need to pay a non-zero interest.
The interest rate is used to control the number of created coins.
For instance, raising interest rates makes borrowing stablecoins more expensive---supply decreases.
A project that has launched a system with interest rates for both deposits and loans is \emph{Maker}~\cite{dai} with its token \emph{Dai}.
Since December 2017, it has deviated from a \SI{5}{\percent} band around the 1 USD peg, with a single day at 0.94~USD.

\subsection{Currency interventions}

Currency interventions are a technique for a direct money supply adjustment.
Here, an abstract monetary actor in the form of multiple persons and/or trading bots, intervenes in currency markets by buying and selling coins in exchange for the currency to which the stablecoin is pegged.
When demand increases, coins are created and sold on the market for reserves.
This increases the money supply to match the increased demand and subsequently normalize the price. 
In the opposite situation, when demand decreases, coins have to be bought back, decreasing supply and therefore stabilizing the price again.
In contrast to collateralization where market participants are incentivized to stabilize the price through the backing with collateral, currency interventions require active intervention by some actor related to the stablecoin.
Naturally, the purchase of coins requires that the monetary actor has currency reserves that can be spent on the market.
Once the reserves are depleted, the exchange rate is governed by market forces, which can lead to a drastic change in the price and damage trust.

\subsection{Open market operations}

Open Market Operations (OMO) can be seen as a generalization of the currency interventions technique. 
A monetary actor manually or (semi-)automatically purchases external assets and pays them with newly minted money which increases the money supply.
The system contracts supply by selling the assets back to the market\footnote{
We differentiate between currency interventions and OMO to highlight the specific feature of the former. 
Buying or selling the targeted currency against stable coins implies a more effective impact on the mutual exchange rate as the supplies of the targeted currency and stable coin move in the opposite direction simultaneously.
}.
For instance, if the Fed buys U.S. Treasury Securities on the open market, it effectively increases the supply of dollars.
Selling these securities back to the market allows to decrease the supply again.
A number of stablecoins implicitly or explicitly consider replicating this technique.
The proposed designs, however, ignore certain safeguards often used by national central banks.

The most important of these safeguards are \emph{eligibility} and \emph{reversibility}.
Eligibility demands that only highly secure and liquid third-party assets can built central bank reserves (compare \cite{ecb_eligibility1}, \cite{ecb_eligibility2}, \cite{ecb_eligibility3} or \cite{bis2009collateral}, \cite{cheun2009collateral}). 
This ensures that supply can be decreased in the future by selling the assets.
The higher their price, the more money supply can be absorbed.
Reversibility requires, that OMO is automatically reversed after a predetermined period.
This ensures that, by default, supply increases only short term.

To highlight negligence of the above safeguard principles, we differentiated between three sub-categories:

\emph{Standard OMO} classifies OMO implementations satisfying the eligibility and reversibility safeguards.
None of the proposed techniques in current projects can be classified as such.   

\emph{Proxy OMO} violates at least one of the safeguards.
Proposals in projects like \emph{Celo}~\cite{celo} or \emph{Augmint}~\cite{augmints} are examples.

\emph{Self-tokenizing OMO} decreases supply not by selling external assets, but other assets generated within their own ecosystem.
Examples are \emph{Basecoin}~\cite{basecoin}, \emph{Carbon}~\cite{carbon} and \emph{Fragments}~\cite{fragments}.
All these projects target a 1-to-1 relationship between their stablecoin and the USD. 
To decrease the money supply, the projects propose mechanisms that create special-purpose tokens that are sold for stablecoins which are then destroyed by the system.
In theory, with such a design, supply can be decreased to any desired level.
This is different from standard and proxy OMO that are restricted by the available reserves of external assets.
In practice, many open questions remain (addressed in Sec.~\ref{subsec:subsec:stab_techniques_focalpoints}).

\subsection{Use of dynamic block rewards and dynamically burned transaction fees}

Instead of a pre-defined change in the money supply as in \emph{Bitcoin}\cite{bitcoin} or \emph{Ethereum}\cite{ethereum}, the mining reward can depend on the current state of the system.
If supply needs to be increased this can be done by increasing the mining reward. 
Since a very low or even negative block reward is not practical, this technique can only increase supply.
Furthermore, increasing the mining reward is equivalent to ``printing'' money since currency is issued without any backing.
Note that a variable block reward leads to variability in the hash rate due to varying incentives to increase/decrease mining power---elevating the risk of double-spending attacks~\cite{DBLP:journals/corr/Rosenfeld14}.
To provide a way to decrease supply, some projects suggest \emph{dynamically burned transaction fees}, \ie, a part of transactions fees is not given to miners but burned instead.
Hence, the possibility to decrease supply is limited by the total volume of fees over a period of time.

\section{Stabilization techniques: Discussion}
\label{subsec:subsec:stab_techniques_focalpoints}

In the preceding section we presented the stabilization concepts underlying the surveyed stablecoins.
We purposefully disconnected the in-depth description of the techniques from the discussion of their merits and drawbacks, which is the center of this section.

\subsection{Tokenization of collateral}

While in theory collateralization in itself provides an elegant way to link money supply and demand through the action of market participants, it exhibits several risks and limitations that render this technique less reliable in practice.
Direct collateralization, due the to link to fiat currencies or traditional assets, requires a trusted third party that manages funds, assets and the correct issuance of tokens.
While the resulting counterparty risk can be remedied to some degree by, \eg, escrow accounts and diversified banking partners, the necessity of a trusted third party remains as a major limitation.

Proxy collateralization could help to avoid above risks, since the collateral can be another cryptocurrency (\eg, \emph{Bitcoin}\cite{bitcoin} or \emph{Ether}\cite{ethereum}).
However, even if the counterparty risk can be eliminated, the requirement of a trusted price feed gives rise to the oracle problem (cf. Sec.~\ref{sec:dec_trust}).
Furthermore, if the collateral's value fluctuates (as it is the case for cryptocurrencies), price risk of the collateral has to be mitigated.
Margin calls are often cited as a remedy for collateral risk, however, margin calls require the assumption that markets for the collateral asset are liquid and large enough to allow for timely provision or absorption of collateral.
This assumption might become an issue for young stablecoin projects or if expectations on the future development of the collateral are dire.

As a subform of proxy collateralization, self collateralization exhibits the same risks.
Moreover, it suffers from additional systematic risk between the collateral and the stablecoin, as the value of the collateral is often a function of the future expected demand on the stablecoin.

\subsection{Interest rates on deposits and loans}

As discussed in Sec.~\ref{sec:stab_techniques}, interest rates on deposits can reduce the money supply only temporarily and thus should be coupled with other techniques.
When it comes to loans, an interesting question is whether under-collateralized loans can be implemented at all.
This is the case in the regular economy, where wealth or future income can be used as collateral.
We argue that this is impossible in permissionless cryptocurrencies due to the lack of strong identities and the resulting vulnerability to Sybil attacks~\cite{douceur2002sybil}.
If under-collateralized loans were implemented, rational actors would spawn multiple fake identities to obtain loans and free money.

\begin{theorem}
  In a permissionless setting without strong identities, under-collateralized loans enable arbitrage to the point where only fully collateralized loans are available.
  \begin{proof}
    Let $L$ be a loan that can be taken by depositing an amount of collateral $C$, with $p_L$ and $p_C$ denoting the respective prices.
    In an under-collateralized setting $p_C < p_L$.
    A rational agent would seize the arbitrage opportunity, spend $p_C$ on collateral and receive a loan with value $p_L = p_C + \epsilon$.
    The loan can be used to purchase more collateral and create more loans, generating a profit of $\epsilon_i$ in each step $i$, until the arbitrage opportunity closes due to increased collateral demand, \ie $p_C \nless p_L$.
    Since there are no identities, the agent can refuse to repay the loans he has taken without any risk, locking the collateral forever and still generating a profit of $\sum\epsilon_i$.
  \end{proof}
\end{theorem}

Even with smart contracts that enforce payments and interest rates, the lack of strong identities makes it easy to simply ``exit-scam'' the system, \ie, to generate a new debt-free identity and start over without negative consequences.

\subsection{Currency Interventions}
For currency interventions, the ability to maintain a peg during falling prices is limited by the amount of available reserves and the monetary authorities' commitment to make use of them.
Once the reserves are depleted, the exchange rate is governed by market forces, which can lead to a drastic change in the exchange rate.
The usage of currency interventions can, under certain assumptions, increase the vulnerability to speculative attacks (cf. Sec.~\ref{subsec:era_speculativeattacks}).
The interplay of full transparency of the system and gameability of the intended interventions is an interesting open question.

\subsection{Open market operations}
The negligence of safeguards by techniques classified as proxy OMO is no triviality. High quality (eligibility) of the assets seized by the cryptocurrency system prevents erosion of its reserves that can be used to buy back outstanding currency units.
The programmatic reversal of open market arrangements ensures that a long term expansion of the money supply is not possible without manually overriding the default policy.
While reviewed projects only allow using cryptocurrencies with a relatively long track record (\emph{Bitcoin}\cite{bitcoin}, \emph{Ether}\cite{ethereum}), reversibility has not been proposed yet.

In self-tokenizing OMO not reserves but special-purpose tokens are sold against currency units.
While the designs of the tokens vary, all provide some form of success-related monetary incentive that is payed out if a certain target price for the stablecoin is achieved.
The incentive is paid out in form of newly minted money supply.
The lower the probability of success, the higher the necessary incentives. 
Risk is either remunerated by higher relative ownership of future money growth or by a promised absolute increase of future minted money.
Excessive use thus may either lead to reduced ownership in risk remuneration or to an uncontrolled increase of promises of future remuneration in the money supply.
Similar to currency interventions, OMO setups are vulnerable to speculative attacks (c.f. Sec.~\ref{subsec:era_speculativeattacks}).
Lastly, the technique can decrease supply only in the short run, as long as token remuneration promises are not retracted.

\subsection{Dynamic mining reward}
\label{subsec:subsec:stab_techniques_focalpoints:mining}
Mining is a vital function of most cryptocurrency systems.
The goal of making the money supply dynamic should be subordinated to the security and usability of the financial system.
Low block rewards or high difficulty, \eg, in phases of stagnating demand for the currency, would lead to less incentive for miners to process transactions.
This would necessarily lead to lower transaction throughput, which would not only reduce the liquidity of the coin, but also increase the risk of double-spending attacks.
Moreover, as this technique cannot be used to reduce the money supply, it should be coupled with other instruments.

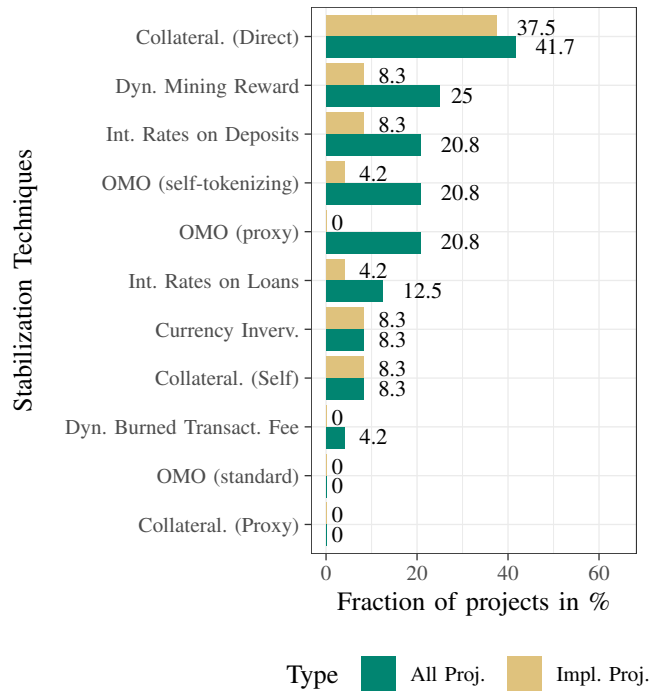
\begin{figure}[h]
\centering
\begin{tikzpicture}[x=1pt,y=1pt]
\definecolor{fillColor}{RGB}{255,255,255}
\path[use as bounding box,fill=fillColor,fill opacity=0.00] (0,0) rectangle (245.72,274.63);
\begin{scope}
\path[clip] (  0.00,  0.00) rectangle (245.72,274.63);
\definecolor{drawColor}{RGB}{255,255,255}
\definecolor{fillColor}{RGB}{255,255,255}

\path[draw=drawColor,line width= 0.5pt,line join=round,line cap=round,fill=fillColor] ( -0.00,  0.00) rectangle (245.72,274.63);
\end{scope}
\begin{scope}
\path[clip] (117.67, 62.74) rectangle (240.72,269.63);
\definecolor{fillColor}{RGB}{255,255,255}

\path[fill=fillColor] (117.67, 62.74) rectangle (240.72,269.63);
\definecolor{drawColor}{gray}{0.92}

\path[draw=drawColor,line width= 0.3pt,line join=round] (140.48, 62.74) --
  (140.48,269.63);

\path[draw=drawColor,line width= 0.3pt,line join=round] (174.89, 62.74) --
  (174.89,269.63);

\path[draw=drawColor,line width= 0.3pt,line join=round] (209.31, 62.74) --
  (209.31,269.63);

\path[draw=drawColor,line width= 0.5pt,line join=round] (117.67, 73.82) --
  (240.72, 73.82);

\path[draw=drawColor,line width= 0.5pt,line join=round] (117.67, 92.29) --
  (240.72, 92.29);

\path[draw=drawColor,line width= 0.5pt,line join=round] (117.67,110.77) --
  (240.72,110.77);

\path[draw=drawColor,line width= 0.5pt,line join=round] (117.67,129.24) --
  (240.72,129.24);

\path[draw=drawColor,line width= 0.5pt,line join=round] (117.67,147.71) --
  (240.72,147.71);

\path[draw=drawColor,line width= 0.5pt,line join=round] (117.67,166.18) --
  (240.72,166.18);

\path[draw=drawColor,line width= 0.5pt,line join=round] (117.67,184.65) --
  (240.72,184.65);

\path[draw=drawColor,line width= 0.5pt,line join=round] (117.67,203.13) --
  (240.72,203.13);

\path[draw=drawColor,line width= 0.5pt,line join=round] (117.67,221.60) --
  (240.72,221.60);

\path[draw=drawColor,line width= 0.5pt,line join=round] (117.67,240.07) --
  (240.72,240.07);

\path[draw=drawColor,line width= 0.5pt,line join=round] (117.67,258.54) --
  (240.72,258.54);

\path[draw=drawColor,line width= 0.5pt,line join=round] (123.27, 62.74) --
  (123.27,269.63);

\path[draw=drawColor,line width= 0.5pt,line join=round] (157.69, 62.74) --
  (157.69,269.63);

\path[draw=drawColor,line width= 0.5pt,line join=round] (192.10, 62.74) --
  (192.10,269.63);

\path[draw=drawColor,line width= 0.5pt,line join=round] (226.52, 62.74) --
  (226.52,269.63);
\definecolor{fillColor}{RGB}{223,194,125}

\path[fill=fillColor] (123.27, 73.82) rectangle (123.27, 82.13);
\definecolor{fillColor}{RGB}{1,133,113}

\path[fill=fillColor] (123.27, 65.51) rectangle (123.27, 73.82);
\definecolor{fillColor}{RGB}{223,194,125}

\path[fill=fillColor] (123.27, 92.29) rectangle (123.27,100.61);
\definecolor{fillColor}{RGB}{1,133,113}

\path[fill=fillColor] (123.27, 83.98) rectangle (123.27, 92.29);
\definecolor{fillColor}{RGB}{223,194,125}

\path[fill=fillColor] (123.27,110.77) rectangle (123.27,119.08);
\definecolor{fillColor}{RGB}{1,133,113}

\path[fill=fillColor] (123.27,102.45) rectangle (130.44,110.77);
\definecolor{fillColor}{RGB}{223,194,125}

\path[fill=fillColor] (123.27,129.24) rectangle (137.60,137.55);
\definecolor{fillColor}{RGB}{1,133,113}

\path[fill=fillColor] (123.27,120.93) rectangle (137.60,129.24);
\definecolor{fillColor}{RGB}{223,194,125}

\path[fill=fillColor] (123.27,147.71) rectangle (137.60,156.02);
\definecolor{fillColor}{RGB}{1,133,113}

\path[fill=fillColor] (123.27,139.40) rectangle (137.60,147.71);
\definecolor{fillColor}{RGB}{223,194,125}

\path[fill=fillColor] (123.27,166.18) rectangle (130.44,174.49);
\definecolor{fillColor}{RGB}{1,133,113}

\path[fill=fillColor] (123.27,157.87) rectangle (144.78,166.18);
\definecolor{fillColor}{RGB}{223,194,125}

\path[fill=fillColor] (123.27,184.65) rectangle (123.27,192.97);
\definecolor{fillColor}{RGB}{1,133,113}

\path[fill=fillColor] (123.27,176.34) rectangle (159.11,184.65);
\definecolor{fillColor}{RGB}{223,194,125}

\path[fill=fillColor] (123.27,203.13) rectangle (130.44,211.44);
\definecolor{fillColor}{RGB}{1,133,113}

\path[fill=fillColor] (123.27,194.81) rectangle (159.11,203.13);
\definecolor{fillColor}{RGB}{223,194,125}

\path[fill=fillColor] (123.27,221.60) rectangle (137.60,229.91);
\definecolor{fillColor}{RGB}{1,133,113}

\path[fill=fillColor] (123.27,213.29) rectangle (159.11,221.60);
\definecolor{fillColor}{RGB}{223,194,125}

\path[fill=fillColor] (123.27,240.07) rectangle (137.60,248.38);
\definecolor{fillColor}{RGB}{1,133,113}

\path[fill=fillColor] (123.27,231.76) rectangle (166.29,240.07);
\definecolor{fillColor}{RGB}{223,194,125}

\path[fill=fillColor] (123.27,258.54) rectangle (187.80,266.86);
\definecolor{fillColor}{RGB}{1,133,113}

\path[fill=fillColor] (123.27,250.23) rectangle (194.98,258.54);
\definecolor{drawColor}{RGB}{0,0,0}

\node[text=drawColor,anchor=base west,inner sep=0pt, outer sep=0pt, scale=  0.85] at (125.40, 74.58) {0};

\node[text=drawColor,anchor=base west,inner sep=0pt, outer sep=0pt, scale=  0.85] at (125.40, 67.19) {0};

\node[text=drawColor,anchor=base west,inner sep=0pt, outer sep=0pt, scale=  0.85] at (125.40, 93.05) {0};

\node[text=drawColor,anchor=base west,inner sep=0pt, outer sep=0pt, scale=  0.85] at (125.40, 85.66) {0};

\node[text=drawColor,anchor=base west,inner sep=0pt, outer sep=0pt, scale=  0.85] at (125.40,111.52) {0};

\node[text=drawColor,anchor=base west,inner sep=0pt, outer sep=0pt, scale=  0.85] at (135.90,104.13) {4.2};

\node[text=drawColor,anchor=base west,inner sep=0pt, outer sep=0pt, scale=  0.85] at (143.05,129.99) {8.3};

\node[text=drawColor,anchor=base west,inner sep=0pt, outer sep=0pt, scale=  0.85] at (143.05,122.60) {8.3};

\node[text=drawColor,anchor=base west,inner sep=0pt, outer sep=0pt, scale=  0.85] at (143.05,148.47) {8.3};

\node[text=drawColor,anchor=base west,inner sep=0pt, outer sep=0pt, scale=  0.85] at (143.05,141.08) {8.3};

\node[text=drawColor,anchor=base west,inner sep=0pt, outer sep=0pt, scale=  0.85] at (135.90,166.94) {4.2};

\node[text=drawColor,anchor=base west,inner sep=0pt, outer sep=0pt, scale=  0.85] at (152.36,159.55) {12.5};

\node[text=drawColor,anchor=base west,inner sep=0pt, outer sep=0pt, scale=  0.85] at (125.40,185.41) {0};

\node[text=drawColor,anchor=base west,inner sep=0pt, outer sep=0pt, scale=  0.85] at (166.70,178.02) {20.8};

\node[text=drawColor,anchor=base west,inner sep=0pt, outer sep=0pt, scale=  0.85] at (135.90,203.88) {4.2};

\node[text=drawColor,anchor=base west,inner sep=0pt, outer sep=0pt, scale=  0.85] at (166.70,196.49) {20.8};

\node[text=drawColor,anchor=base west,inner sep=0pt, outer sep=0pt, scale=  0.85] at (143.05,222.35) {8.3};

\node[text=drawColor,anchor=base west,inner sep=0pt, outer sep=0pt, scale=  0.85] at (166.70,214.96) {20.8};

\node[text=drawColor,anchor=base west,inner sep=0pt, outer sep=0pt, scale=  0.85] at (143.05,240.83) {8.3};

\node[text=drawColor,anchor=base west,inner sep=0pt, outer sep=0pt, scale=  0.85] at (170.56,233.44) {25};

\node[text=drawColor,anchor=base west,inner sep=0pt, outer sep=0pt, scale=  0.85] at (195.39,259.30) {37.5};

\node[text=drawColor,anchor=base west,inner sep=0pt, outer sep=0pt, scale=  0.85] at (202.56,251.91) {41.7};
\definecolor{drawColor}{gray}{0.20}

\path[draw=drawColor,line width= 0.5pt,line join=round,line cap=round] (117.67, 62.74) rectangle (240.72,269.63);
\end{scope}
\begin{scope}
\path[clip] (  0.00,  0.00) rectangle (245.72,274.63);
\definecolor{drawColor}{gray}{0.30}

\node[text=drawColor,anchor=base east,inner sep=0pt, outer sep=0pt, scale=  0.80] at (113.17, 71.07) {Collateral. (Proxy)};

\node[text=drawColor,anchor=base east,inner sep=0pt, outer sep=0pt, scale=  0.80] at (113.17, 89.54) {OMO (standard)};

\node[text=drawColor,anchor=base east,inner sep=0pt, outer sep=0pt, scale=  0.80] at (113.17,108.01) {Dyn. Burned Transact. Fee};

\node[text=drawColor,anchor=base east,inner sep=0pt, outer sep=0pt, scale=  0.80] at (113.17,126.48) {Collateral. (Self)};

\node[text=drawColor,anchor=base east,inner sep=0pt, outer sep=0pt, scale=  0.80] at (113.17,144.96) {Currency Inverv.};

\node[text=drawColor,anchor=base east,inner sep=0pt, outer sep=0pt, scale=  0.80] at (113.17,163.43) {Int. Rates on Loans};

\node[text=drawColor,anchor=base east,inner sep=0pt, outer sep=0pt, scale=  0.80] at (113.17,181.90) {OMO (proxy)};

\node[text=drawColor,anchor=base east,inner sep=0pt, outer sep=0pt, scale=  0.80] at (113.17,200.37) {OMO (self-tokenizing)};

\node[text=drawColor,anchor=base east,inner sep=0pt, outer sep=0pt, scale=  0.80] at (113.17,218.84) {Int. Rates on Deposits};

\node[text=drawColor,anchor=base east,inner sep=0pt, outer sep=0pt, scale=  0.80] at (113.17,237.32) {Dyn. Mining Reward};

\node[text=drawColor,anchor=base east,inner sep=0pt, outer sep=0pt, scale=  0.80] at (113.17,255.79) {Collateral. (Direct)};
\end{scope}
\begin{scope}
\path[clip] (  0.00,  0.00) rectangle (245.72,274.63);
\definecolor{drawColor}{gray}{0.20}

\path[draw=drawColor,line width= 0.5pt,line join=round] (115.17, 73.82) --
  (117.67, 73.82);

\path[draw=drawColor,line width= 0.5pt,line join=round] (115.17, 92.29) --
  (117.67, 92.29);

\path[draw=drawColor,line width= 0.5pt,line join=round] (115.17,110.77) --
  (117.67,110.77);

\path[draw=drawColor,line width= 0.5pt,line join=round] (115.17,129.24) --
  (117.67,129.24);

\path[draw=drawColor,line width= 0.5pt,line join=round] (115.17,147.71) --
  (117.67,147.71);

\path[draw=drawColor,line width= 0.5pt,line join=round] (115.17,166.18) --
  (117.67,166.18);

\path[draw=drawColor,line width= 0.5pt,line join=round] (115.17,184.65) --
  (117.67,184.65);

\path[draw=drawColor,line width= 0.5pt,line join=round] (115.17,203.13) --
  (117.67,203.13);

\path[draw=drawColor,line width= 0.5pt,line join=round] (115.17,221.60) --
  (117.67,221.60);

\path[draw=drawColor,line width= 0.5pt,line join=round] (115.17,240.07) --
  (117.67,240.07);

\path[draw=drawColor,line width= 0.5pt,line join=round] (115.17,258.54) --
  (117.67,258.54);
\end{scope}
\begin{scope}
\path[clip] (  0.00,  0.00) rectangle (245.72,274.63);
\definecolor{drawColor}{gray}{0.20}

\path[draw=drawColor,line width= 0.5pt,line join=round] (123.27, 60.24) --
  (123.27, 62.74);

\path[draw=drawColor,line width= 0.5pt,line join=round] (157.69, 60.24) --
  (157.69, 62.74);

\path[draw=drawColor,line width= 0.5pt,line join=round] (192.10, 60.24) --
  (192.10, 62.74);

\path[draw=drawColor,line width= 0.5pt,line join=round] (226.52, 60.24) --
  (226.52, 62.74);
\end{scope}
\begin{scope}
\path[clip] (  0.00,  0.00) rectangle (245.72,274.63);
\definecolor{drawColor}{gray}{0.30}

\node[text=drawColor,anchor=base,inner sep=0pt, outer sep=0pt, scale=  0.80] at (123.27, 52.73) {0};

\node[text=drawColor,anchor=base,inner sep=0pt, outer sep=0pt, scale=  0.80] at (157.69, 52.73) {20};

\node[text=drawColor,anchor=base,inner sep=0pt, outer sep=0pt, scale=  0.80] at (192.10, 52.73) {40};

\node[text=drawColor,anchor=base,inner sep=0pt, outer sep=0pt, scale=  0.80] at (226.52, 52.73) {60};
\end{scope}
\begin{scope}
\path[clip] (  0.00,  0.00) rectangle (245.72,274.63);
\definecolor{drawColor}{RGB}{0,0,0}

\node[text=drawColor,anchor=base,inner sep=0pt, outer sep=0pt, scale=  1.00] at (179.20, 41.40) {Fraction of projects in \%};
\end{scope}
\begin{scope}
\path[clip] (  0.00,  0.00) rectangle (245.72,274.63);
\definecolor{drawColor}{RGB}{0,0,0}

\node[text=drawColor,rotate= 90.00,anchor=base,inner sep=0pt, outer sep=0pt, scale=  1.00] at ( 11.89,166.18) {Stabilization Techniques};
\end{scope}
\begin{scope}
\path[clip] (  0.00,  0.00) rectangle (245.72,274.63);
\definecolor{fillColor}{RGB}{255,255,255}

\path[fill=fillColor] (103.26,  5.00) rectangle (255.13, 29.45);
\end{scope}
\begin{scope}
\path[clip] (  0.00,  0.00) rectangle (245.72,274.63);
\definecolor{drawColor}{RGB}{0,0,0}

\node[text=drawColor,anchor=base west,inner sep=0pt, outer sep=0pt, scale=  1.00] at (108.26, 13.78) {Type};
\end{scope}
\begin{scope}
\path[clip] (  0.00,  0.00) rectangle (245.72,274.63);
\definecolor{fillColor}{RGB}{255,255,255}

\path[fill=fillColor] (135.75, 10.00) rectangle (150.21, 24.45);
\end{scope}
\begin{scope}
\path[clip] (  0.00,  0.00) rectangle (245.72,274.63);
\definecolor{fillColor}{RGB}{1,133,113}

\path[fill=fillColor] (136.46, 10.71) rectangle (149.50, 23.74);
\end{scope}
\begin{scope}
\path[clip] (  0.00,  0.00) rectangle (245.72,274.63);
\definecolor{fillColor}{RGB}{255,255,255}

\path[fill=fillColor] (191.00, 10.00) rectangle (205.45, 24.45);
\end{scope}
\begin{scope}
\path[clip] (  0.00,  0.00) rectangle (245.72,274.63);
\definecolor{fillColor}{RGB}{223,194,125}

\path[fill=fillColor] (191.71, 10.71) rectangle (204.74, 23.74);
\end{scope}
\begin{scope}
\path[clip] (  0.00,  0.00) rectangle (245.72,274.63);
\definecolor{drawColor}{RGB}{0,0,0}

\node[text=drawColor,anchor=base west,inner sep=0pt, outer sep=0pt, scale=  0.80] at (155.21, 14.47) {All Proj.};
\end{scope}
\begin{scope}
\path[clip] (  0.00,  0.00) rectangle (245.72,274.63);
\definecolor{drawColor}{RGB}{0,0,0}

\node[text=drawColor,anchor=base west,inner sep=0pt, outer sep=0pt, scale=  0.80] at (210.45, 14.47) {Impl. Proj.};
\end{scope}
\end{tikzpicture}

  \caption{Planned and implemented stabilization techniques.}
  \label{fig:stabtechs_crypto}
\end{figure}

\subsection{Classification results and blank spots}
\label{subsec:stab_techniques_blankspots}
Fig.~\ref{fig:stabtechs_crypto} shows a full list of techniques discussed in this section as well as their prevalence of adoption in stablecoins projects, distinguished by planned and implemented.

The most popular technique according to our observations is direct collateralization. 
It is followed by the use of dynamic mining reward, interest rates on deposits and self-tokenizing OMO. None of these methods can permanently decrease money supply. 
Current stablecoin projects plan to launch primarily solutions which either require the participation of a trusted third-party or are focused on techniques that can decrease supply only temporarily and consequently are not sustainable in the long term.

Interest rates on loans, currency interventions, standard OMO and maybe even proxy OMO might be useful techniques as they allow for decreasing the money supply permanently.
However, exactly those have been worked on to a lower degree: although well over \SI{40}{\percent} of projects plan some form of OMO, only around \SI{4}{\percent} implemented their proposed setup.
Note that established monetary policy standards find little acknowledgment---no project implemented the requirements of standard OMO, although other types of OMO are introduced.
As there is little practical experience yet, risks and potentials of these techniques are hard to assess. 

But also for less complex approaches there are blank spots.
None of the reviewed projects has implemented dynamically burned transaction fees or proxy collateralization.

\section{Exchange rate regimes}
\label{sec:era}
So far we implicitly interpreted ``stability'' as stabilizing the price of each stablecoin to \emph{exactly} 1 EUR or 1 USD.
While this seems to be an intuitive approach, other so called \emph{exchange rate regimes} are possible.

\begin{figure}
\centering
\begin{tikzpicture}[x=1pt,y=1pt]
\definecolor{fillColor}{RGB}{255,255,255}
\path[use as bounding box,fill=fillColor,fill opacity=0.00] (0,0) rectangle (245.72,216.81);
\begin{scope}
\path[clip] (  0.00,  0.00) rectangle (245.72,216.81);
\definecolor{drawColor}{RGB}{255,255,255}
\definecolor{fillColor}{RGB}{255,255,255}

\path[draw=drawColor,line width= 0.5pt,line join=round,line cap=round,fill=fillColor] (  0.00,  0.00) rectangle (245.72,216.81);
\end{scope}
\begin{scope}
\path[clip] ( 69.15, 62.74) rectangle (240.72,211.81);
\definecolor{fillColor}{RGB}{255,255,255}

\path[fill=fillColor] ( 69.15, 62.74) rectangle (240.72,211.81);
\definecolor{drawColor}{gray}{0.92}

\path[draw=drawColor,line width= 0.3pt,line join=round] ( 96.45, 62.74) --
  ( 96.45,211.81);

\path[draw=drawColor,line width= 0.3pt,line join=round] (135.44, 62.74) --
  (135.44,211.81);

\path[draw=drawColor,line width= 0.3pt,line join=round] (174.43, 62.74) --
  (174.43,211.81);

\path[draw=drawColor,line width= 0.3pt,line join=round] (213.42, 62.74) --
  (213.42,211.81);

\path[draw=drawColor,line width= 0.5pt,line join=round] ( 69.15, 79.94) --
  (240.72, 79.94);

\path[draw=drawColor,line width= 0.5pt,line join=round] ( 69.15,108.61) --
  (240.72,108.61);

\path[draw=drawColor,line width= 0.5pt,line join=round] ( 69.15,137.27) --
  (240.72,137.27);

\path[draw=drawColor,line width= 0.5pt,line join=round] ( 69.15,165.94) --
  (240.72,165.94);

\path[draw=drawColor,line width= 0.5pt,line join=round] ( 69.15,194.61) --
  (240.72,194.61);

\path[draw=drawColor,line width= 0.5pt,line join=round] ( 76.95, 62.74) --
  ( 76.95,211.81);

\path[draw=drawColor,line width= 0.5pt,line join=round] (115.94, 62.74) --
  (115.94,211.81);

\path[draw=drawColor,line width= 0.5pt,line join=round] (154.94, 62.74) --
  (154.94,211.81);

\path[draw=drawColor,line width= 0.5pt,line join=round] (193.93, 62.74) --
  (193.93,211.81);

\path[draw=drawColor,line width= 0.5pt,line join=round] (232.92, 62.74) --
  (232.92,211.81);
\definecolor{fillColor}{RGB}{166,97,26}

\path[fill=fillColor] ( 76.95, 84.24) rectangle ( 93.17, 92.84);
\definecolor{fillColor}{RGB}{223,194,125}

\path[fill=fillColor] ( 76.95, 75.64) rectangle ( 76.95, 84.24);
\definecolor{fillColor}{RGB}{1,133,113}

\path[fill=fillColor] ( 76.95, 67.04) rectangle ( 76.95, 75.64);
\definecolor{fillColor}{RGB}{166,97,26}

\path[fill=fillColor] ( 76.95,112.91) rectangle (102.06,121.51);
\definecolor{fillColor}{RGB}{223,194,125}

\path[fill=fillColor] ( 76.95,104.31) rectangle ( 83.45,112.91);
\definecolor{fillColor}{RGB}{1,133,113}

\path[fill=fillColor] ( 76.95, 95.71) rectangle ( 83.45,104.31);
\definecolor{fillColor}{RGB}{166,97,26}

\path[fill=fillColor] ( 76.95,141.57) rectangle (109.39,150.17);
\definecolor{fillColor}{RGB}{223,194,125}

\path[fill=fillColor] ( 76.95,132.97) rectangle ( 76.95,141.57);
\definecolor{fillColor}{RGB}{1,133,113}

\path[fill=fillColor] ( 76.95,124.37) rectangle ( 83.45,132.97);
\definecolor{fillColor}{RGB}{166,97,26}

\path[fill=fillColor] ( 76.95,170.24) rectangle ( 97.23,178.84);
\definecolor{fillColor}{RGB}{223,194,125}

\path[fill=fillColor] ( 76.95,161.64) rectangle (135.44,170.24);
\definecolor{fillColor}{RGB}{1,133,113}

\path[fill=fillColor] ( 76.95,153.04) rectangle (141.94,161.64);
\definecolor{fillColor}{RGB}{166,97,26}

\path[fill=fillColor] ( 76.95,198.91) rectangle (138.71,207.51);
\definecolor{fillColor}{RGB}{223,194,125}

\path[fill=fillColor] ( 76.95,190.31) rectangle (109.44,198.91);
\definecolor{fillColor}{RGB}{1,133,113}

\path[fill=fillColor] ( 76.95,181.71) rectangle (154.94,190.31);
\definecolor{drawColor}{RGB}{0,0,0}

\node[text=drawColor,anchor=base west,inner sep=0pt, outer sep=0pt, scale=  0.85] at ( 93.93, 84.64) {10.4};

\node[text=drawColor,anchor=base west,inner sep=0pt, outer sep=0pt, scale=  0.85] at ( 77.16, 77.00) {0};

\node[text=drawColor,anchor=base west,inner sep=0pt, outer sep=0pt, scale=  0.85] at ( 77.16, 69.36) {0};

\node[text=drawColor,anchor=base west,inner sep=0pt, outer sep=0pt, scale=  0.85] at (102.82,113.31) {16.1};

\node[text=drawColor,anchor=base west,inner sep=0pt, outer sep=0pt, scale=  0.85] at ( 84.00,105.67) {4.2};

\node[text=drawColor,anchor=base west,inner sep=0pt, outer sep=0pt, scale=  0.85] at ( 84.00, 98.02) {4.2};

\node[text=drawColor,anchor=base west,inner sep=0pt, outer sep=0pt, scale=  0.85] at (110.15,141.98) {20.8};

\node[text=drawColor,anchor=base west,inner sep=0pt, outer sep=0pt, scale=  0.85] at ( 77.16,134.33) {0};

\node[text=drawColor,anchor=base west,inner sep=0pt, outer sep=0pt, scale=  0.85] at ( 84.00,126.69) {4.2};

\node[text=drawColor,anchor=base west,inner sep=0pt, outer sep=0pt, scale=  0.85] at ( 97.65,170.65) {13};

\node[text=drawColor,anchor=base west,inner sep=0pt, outer sep=0pt, scale=  0.85] at (136.20,163.00) {37.5};

\node[text=drawColor,anchor=base west,inner sep=0pt, outer sep=0pt, scale=  0.85] at (142.70,155.36) {41.7};

\node[text=drawColor,anchor=base west,inner sep=0pt, outer sep=0pt, scale=  0.85] at (139.47,199.31) {39.6};

\node[text=drawColor,anchor=base west,inner sep=0pt, outer sep=0pt, scale=  0.85] at (110.20,191.67) {20.8};

\node[text=drawColor,anchor=base west,inner sep=0pt, outer sep=0pt, scale=  0.85] at (155.36,184.03) {50};
\definecolor{drawColor}{gray}{0.20}

\path[draw=drawColor,line width= 0.5pt,line join=round,line cap=round] ( 69.15, 62.74) rectangle (240.72,211.81);
\end{scope}
\begin{scope}
\path[clip] (  0.00,  0.00) rectangle (245.72,216.81);
\definecolor{drawColor}{gray}{0.30}

\node[text=drawColor,anchor=base east,inner sep=0pt, outer sep=0pt, scale=  0.80] at ( 64.65, 77.18) {Residual};

\node[text=drawColor,anchor=base east,inner sep=0pt, outer sep=0pt, scale=  0.80] at ( 64.65,105.85) {Float. w. Int.};

\node[text=drawColor,anchor=base east,inner sep=0pt, outer sep=0pt, scale=  0.80] at ( 64.65,134.52) {Free Float};

\node[text=drawColor,anchor=base east,inner sep=0pt, outer sep=0pt, scale=  0.80] at ( 64.65,163.19) {Hard Peg};

\node[text=drawColor,anchor=base east,inner sep=0pt, outer sep=0pt, scale=  0.80] at ( 64.65,191.85) {Soft Peg};
\end{scope}
\begin{scope}
\path[clip] (  0.00,  0.00) rectangle (245.72,216.81);
\definecolor{drawColor}{gray}{0.20}

\path[draw=drawColor,line width= 0.5pt,line join=round] ( 66.65, 79.94) --
  ( 69.15, 79.94);

\path[draw=drawColor,line width= 0.5pt,line join=round] ( 66.65,108.61) --
  ( 69.15,108.61);

\path[draw=drawColor,line width= 0.5pt,line join=round] ( 66.65,137.27) --
  ( 69.15,137.27);

\path[draw=drawColor,line width= 0.5pt,line join=round] ( 66.65,165.94) --
  ( 69.15,165.94);

\path[draw=drawColor,line width= 0.5pt,line join=round] ( 66.65,194.61) --
  ( 69.15,194.61);
\end{scope}
\begin{scope}
\path[clip] (  0.00,  0.00) rectangle (245.72,216.81);
\definecolor{drawColor}{gray}{0.20}

\path[draw=drawColor,line width= 0.5pt,line join=round] ( 76.95, 60.24) --
  ( 76.95, 62.74);

\path[draw=drawColor,line width= 0.5pt,line join=round] (115.94, 60.24) --
  (115.94, 62.74);

\path[draw=drawColor,line width= 0.5pt,line join=round] (154.94, 60.24) --
  (154.94, 62.74);

\path[draw=drawColor,line width= 0.5pt,line join=round] (193.93, 60.24) --
  (193.93, 62.74);

\path[draw=drawColor,line width= 0.5pt,line join=round] (232.92, 60.24) --
  (232.92, 62.74);
\end{scope}
\begin{scope}
\path[clip] (  0.00,  0.00) rectangle (245.72,216.81);
\definecolor{drawColor}{gray}{0.30}

\node[text=drawColor,anchor=base,inner sep=0pt, outer sep=0pt, scale=  0.80] at ( 76.95, 52.73) {0};

\node[text=drawColor,anchor=base,inner sep=0pt, outer sep=0pt, scale=  0.80] at (115.94, 52.73) {25};

\node[text=drawColor,anchor=base,inner sep=0pt, outer sep=0pt, scale=  0.80] at (154.94, 52.73) {50};

\node[text=drawColor,anchor=base,inner sep=0pt, outer sep=0pt, scale=  0.80] at (193.93, 52.73) {75};

\node[text=drawColor,anchor=base,inner sep=0pt, outer sep=0pt, scale=  0.80] at (232.92, 52.73) {100};
\end{scope}
\begin{scope}
\path[clip] (  0.00,  0.00) rectangle (245.72,216.81);
\definecolor{drawColor}{RGB}{0,0,0}

\node[text=drawColor,anchor=base,inner sep=0pt, outer sep=0pt, scale=  1.00] at (154.94, 41.40) {Fraction of projects and countries in \%};
\end{scope}
\begin{scope}
\path[clip] (  0.00,  0.00) rectangle (245.72,216.81);
\definecolor{drawColor}{RGB}{0,0,0}

\node[text=drawColor,rotate= 90.00,anchor=base,inner sep=0pt, outer sep=0pt, scale=  1.00] at ( 11.89,137.27) {Exchange Rate Arrangement};
\end{scope}
\begin{scope}
\path[clip] (  0.00,  0.00) rectangle (245.72,216.81);
\definecolor{fillColor}{RGB}{255,255,255}

\path[fill=fillColor] ( 59.05,  5.00) rectangle (250.82, 29.45);
\end{scope}
\begin{scope}
\path[clip] (  0.00,  0.00) rectangle (245.72,216.81);
\definecolor{drawColor}{RGB}{0,0,0}

\node[text=drawColor,anchor=base west,inner sep=0pt, outer sep=0pt, scale=  1.00] at ( 64.05, 13.78) {Type};
\end{scope}
\begin{scope}
\path[clip] (  0.00,  0.00) rectangle (245.72,216.81);
\definecolor{fillColor}{RGB}{255,255,255}

\path[fill=fillColor] ( 91.55, 10.00) rectangle (106.00, 24.45);
\end{scope}
\begin{scope}
\path[clip] (  0.00,  0.00) rectangle (245.72,216.81);
\definecolor{fillColor}{RGB}{1,133,113}

\path[fill=fillColor] ( 92.26, 10.71) rectangle (105.29, 23.74);
\end{scope}
\begin{scope}
\path[clip] (  0.00,  0.00) rectangle (245.72,216.81);
\definecolor{fillColor}{RGB}{255,255,255}

\path[fill=fillColor] (146.79, 10.00) rectangle (161.25, 24.45);
\end{scope}
\begin{scope}
\path[clip] (  0.00,  0.00) rectangle (245.72,216.81);
\definecolor{fillColor}{RGB}{223,194,125}

\path[fill=fillColor] (147.50, 10.71) rectangle (160.53, 23.74);
\end{scope}
\begin{scope}
\path[clip] (  0.00,  0.00) rectangle (245.72,216.81);
\definecolor{fillColor}{RGB}{255,255,255}

\path[fill=fillColor] (210.93, 10.00) rectangle (225.38, 24.45);
\end{scope}
\begin{scope}
\path[clip] (  0.00,  0.00) rectangle (245.72,216.81);
\definecolor{fillColor}{RGB}{166,97,26}

\path[fill=fillColor] (211.64, 10.71) rectangle (224.67, 23.74);
\end{scope}
\begin{scope}
\path[clip] (  0.00,  0.00) rectangle (245.72,216.81);
\definecolor{drawColor}{RGB}{0,0,0}

\node[text=drawColor,anchor=base west,inner sep=0pt, outer sep=0pt, scale=  0.80] at (111.00, 14.47) {All Proj.};
\end{scope}
\begin{scope}
\path[clip] (  0.00,  0.00) rectangle (245.72,216.81);
\definecolor{drawColor}{RGB}{0,0,0}

\node[text=drawColor,anchor=base west,inner sep=0pt, outer sep=0pt, scale=  0.80] at (166.25, 14.47) {Impl. Proj.};
\end{scope}
\begin{scope}
\path[clip] (  0.00,  0.00) rectangle (245.72,216.81);
\definecolor{drawColor}{RGB}{0,0,0}

\node[text=drawColor,anchor=base west,inner sep=0pt, outer sep=0pt, scale=  0.80] at (230.38, 14.47) {IMF};
\end{scope}
\end{tikzpicture}

  \caption{Exchange rate arrangements: stablecoins and national currencies.}
  \label{fig:ERArrangements_comparison}
\end{figure}
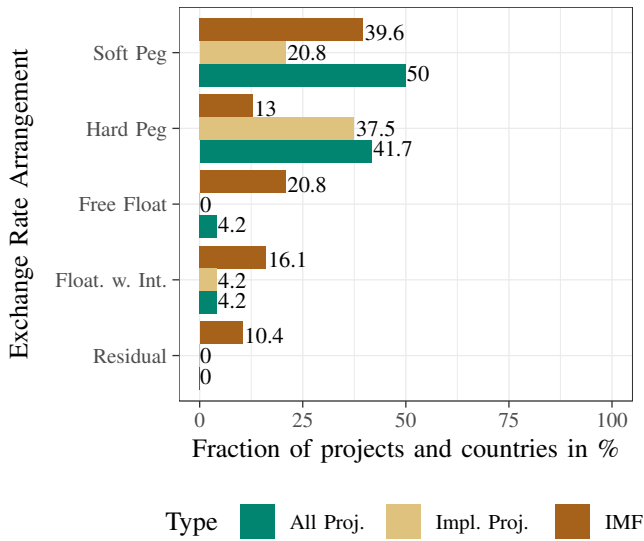

\subsection{Types of exchange rate regimes}
\label{subsec:era_types}
We build upon a taxonomy of the IMF~\cite{international2016exchange} which splits exchange rate regimes into three main types: \emph{hard pegs}, \emph{soft pegs} and \emph{floating regimes}.
Compare Appendix \ref{appendix:taxonomy} for a hierarchical representation of these regimes.

A \emph{hard peg} can come in one of two flavors: arrangements without legal tender and so-called \emph{currency boards}.
In an arrangement without legal tender a country chooses to simply use a well-known foreign currency like the USD instead of issuing their own.\footnote{
  Note that, on a more general level, currencies can also peg against external assets (\eg, gold).
  }
We neglect this case since it is clearly not useful for cryptocurrencies:
it would essentially mean to avoid them altogether.
In a currency board the domestic currency is backed 1:1 (or more) by reserves of the foreign currency~\cite{frankel1999no}.
That is, for every issued unit of domestic currency, there has to be at least one unit of the foreign currency in the reserves.

Different from hard pegs, \emph{soft pegs} are characterized by weaker commitments to a fixed rate, \ie, there does not need to be a 1:1 backing with reserves.
Soft pegs come in a variety of different flavors.
The most important are: conventional pegs, pegs with horizontal bands and crawling pegs.

A \emph{conventional peg} is defined by the level of allowed deviations.
The IMF specifies a maximum fluctuation of \SI{1}{\percent} over a time period of at least six months around the pegged value.
A weaker form is the so-called \emph{peg with horizontal bands}, where the exchange rate is allowed to fluctuate within a pre-announced (wider) range around the pegged value.
These peg types share a common property: the exchange rate is constant over time.
In contrast, \emph{crawling pegs} allow for a gradual adjustment in the exchange rate.

Last but not least, if the exchange rate is \emph{floating}, little to no guarantees are given about the stability of the value.
Instead, the exchange rate is determined by market forces to a large degree and monetary interventions are kept to a minimum.
Due to the fact that free floating can lead to high volatility, some countries intervene aggressively against short term fluctuations (compare Sec.~\ref{sec:stab_techniques}).
This practice is known as ``smoothing'' or \emph{floating with interventions}.

\subsection{Classification results and blank spots}
\label{subsec:era_blankspots}
Fig.~\ref{fig:ERArrangements_comparison} shows a comparison of exchange rate arrangements in stablecoins and traditional central banks.\footnote{
Note that the category ``residual'' refers mainly to countries with frequently changing monetary policy approaches.
}
The data for central banks stems from a study of the IMF in 2016~\cite{international2016exchange}.

The majority of stablecoins, \SI{91.7}{\percent}, commit to achieve some kind of peg.
As in traditional central banking, in cryptocurrencies one has to distinguish between what is announced (de-jure) and the historical exchange rate (de-facto).
De-jure the majority of stablecoins commit themselves to a fixed 1:1 correspondence to the USD in their whitepapers.
\SI{41.7}{\percent} of all projects, tries to enforce this by establishing a currency board and storing the fiat currency pegged to.
Implemented examples include \emph{Tether}~\cite{tether}, \emph{Stasis}~\cite{stasis} and \emph{Trusttoken}~\cite{trusttoken}.

The remainder (\SI{50}{\percent} of all projects), does not implement a currency board and are therefore classified as a soft peg.
Examples include \emph{Maker}~\cite{dai}, \emph{Stasis}~\cite{stably}, \emph{Nubits)}~\cite{nubits}, \emph{Synthetix}~\cite{synthetix} and \emph{Bitshares}~\cite{bitshares}.
Most abstain from explicitly specifying bands and are therefore conventional soft pegs.\footnote{
Some projects mention ``some'' corridor around the peg. As bands are supposed to be predetermined and announced for accountability reasons, we still classify them as conventional pegs.
}

\begin{table}
	\centering
	\caption{Percentage of days for which certain bands around the 1 USD peg are violated.}
  \begin{tabular}{@{\extracolsep{-7pt}} ccccc} 
\\[-1.8ex]\hline 
\hline \\[-1.8ex] 
Projects & \SI{\pm 1}{\percent} & \SI{\pm 5}{\percent} & \SI{\pm 10}{\percent} & \SI{\pm 20}{\percent} \\ 
\hline \\[-1.8ex] 
Tether (USDT) & $12.11$ & $1.05$ & $0$ & $0$ \\ 
Maker (Dai) & $25.25$ & $0.50$ & $0$ & $0$ \\ 
Trusttoken (TrueUSD) & $30.75$ & $0.60$ & $0.60$ & $0$ \\ 
Sythetix (SUSD) & $33.66$ & $3.90$ & $0.98$ & $0$ \\ 
Centre (USD Coin) & $66.10$ & $0$ & $0$ & $0$ \\ 
Stronghold (USDS) & $80.43$ & $8.70$ & $0$ & $0$ \\ 
BitShares (BitUSD) & $70.95$ & $30.23$ & $13.84$ & $4.26$ \\ 
NuBits (Nubits) & $42.22$ & $27.22$ & $26.59$ & $24.07$ \\ 
USC (USC) & $56.25$ & $31.25$ & $31.25$ & $25$ \\ 
\hline \\[-1.8ex] 
\end{tabular} 
    \label{tbl:peg_violations}
\end{table}

Table~\ref{tbl:peg_violations} shows the fraction of daily closing prices violating certain thresholds between the launch of the respective coin and February 7, 2019.
The Table contains the subset of coins that pursue a 1-to-1 peg to the USD.
De-facto, none of the already launched cryptocurrencies meets the demands that would be posed by the IMF for a working conventional peg.
Interestingly, even \emph{Tether}~\cite{tether} and \emph{Trusttoken}~\cite{trusttoken} violate the requirements, despite implementing currency boards.
These fluctuations may stem from uncertainty caused by a perceived lack of transparency and accountability or lower market liquidity.
National currencies in turn tend to use floating arrangements more often.
Although the majority of analyzed stablecoins pursues pegs, concepts for floating arrangements with interventions are also in development, \eg, \emph{MinexCoin}~\cite{minex} proposes interventions to keep daily price changes from exceeding \SI{5}{\percent}.

\subsection{Vulnerabilities to speculative attacks}
\label{subsec:era_speculativeattacks}
The usefulness of soft pegs is disputed in economic literature.
This standpoint is called the \emph{bipolar view}~\cite{fischer2001exchange,williamson2000exchange}, and is broadly supported by mainstream economists~\cite{fischer2001exchange,eichengreen1999exchange,eichengreen1994speculative,crockett1994monetary}.
The bipolar view suggests that there are only two long-term viable options for currency regimes that care for exchange rates: hard pegs or floating with interventions.
Reasons given are short life expectancy of soft pegs and vulnerability to speculative attacks~\cite{fischer2001exchange}.

Speculative attacks on soft pegs are known from traditional central banking~\cite{krugman1979model,obstfeld1996models}, but the threat is equally applicable to cryptocurrencies.
This is especially relevant considering that \SI{50}{\percent} of stablecoin projects plan on using soft pegs.

If the market believes that a fixed exchange rate is not sustainable, investors will start speculating against it to make a profit in the event that it eventually breaks.
To counteract, central banks have to invest resources to defend the peg, which is costly and oftentimes unsuccessful~\cite{obstfeld1995mirage}.
A vivid illustration of unsuccessful peg defense is the Bank of England's attempt to maintain a fixed Great Britain Pound (GBP)-European Currency Unit (ECU) exchange rate during a speculative attack in 1992 lead by the hedge fund ``Quantum''.\footnote{
  ECU was an artificial currency used within the European Monetary System before the introduction of the Euro in 1999~\cite{eichengreen2000ems}.
}

The investors and, subsequently, other market participants followed a simple algorithm:
\begin{enumerate}
  \item Borrow GPB and sell them, at market price, for German Marks (DM); this is called a \emph{short sale}.
  \item When the peg fails and the exchange rate drops, buy back GBP at a cheaper price and return to lender.
\end{enumerate}
The selling of borrowed GBP for DM increases the supply of GBP and reduces the supply of DM.
Due to the fundamental principles of demand and supply, this in turn leads to an appreciation of DM and depreciation of GBP.
To counteract and maintain the target exchange rate, the Bank of England bought the excess GBP on the foreign exchange market in exchange for their DM reserves.
Furthermore, the Bank of England also increased the base interest rate.
Buying the excess supply of GBP aimed at reducing the supply of GBP on the markets, whereas the increase of the interest rate aimed to increase the demand for GBP.
However, after spending 15 billion USD in foreign reserves in only a single day, the Bank of England eventually had to abandon the pegged arrangement~\cite{harmes2002trouble,eichengreen2000ems}.  
The exchange rate on the market dropped, yielding an estimated profit of 1.5 billion USD\footnote{
\url{https://www.forbes.com/sites/steveschaefer/2015/07/07/forbes-flashback-george-soros-british-pound-euro-ecb/4e0e93346131}}.

There are two main sources that make speculative attacks on pegs highly probable~\cite{krugman1979model,obstfeld1996models}:
unsustainably constructed pegs and untrustworthy commitment to defend the peg.
A peg is unsustainable if the central bank lacks sufficient reserves to invest in the case of a speculative attack.
In cryptocurrencies this vulnerability is increased further, since they often have a small market capitalization and little reserves in comparison to traditional financial assets and currencies.
Furthermore, the complete transparency of reserves due to the transparency of the blockchain makes it easy for speculative attackers to validate the success of their strategy~\cite{bhundia2004new}.

Adapting~\cite{krugman1979model} to cryptocurrencies, consider a situation where the natural floating exchange rate would be lower then the peg.
Among others, reasons might be new vulnerabilities or general uncertainty in cryptocurrencies due to regulation.
In both cases, the stablecoin system would need to intervene over longer periods of time, draining its reserves.
Two longer-term outcomes are possible:
(1) the peg holds or (2) the currency finally depreciates when the intervention capabilities are depleted.

Now consider a user of the coin who chooses to \emph{hold} her position.
This user will have no payoff in case (1) and negative payoff in case (2).
Therefore, the expected payoff from holding is negative.
In contrast, if the user \emph{sells} her coins, the sale can be reverted with little cost in case (1) and can avoid a loss in case (2).
Therefore, the payoff for selling is higher than for holding.
Rational market participants will sell their holdings.

The expected payoff of the sell strategy can even be increased through leverage by borrowing coins.\footnote{
  Since efficient credit markets have not yet developed for all cryptocurrencies, the transaction costs to execute speculative attacks might be increased.
}
Speculators might borrow large quantities of stable coins at the pegged price and sell them on the exchanges: 
if the stability system succeeds in defending the peg, speculators can buy back the coins at the peg and revert their positions with little losses. 
If the attack depletes the reserves of the system, the peg can no longer withstand the selling pressure and the exchange rate depreciates and becomes floating.
Attackers can now buy back the stablecoins much cheaper, give back borrowed coins and keep the difference as profit.

\subsection{Peg hard or do not peg at all?}
\label{subsec:peg_hard}
As discussed, the bipolar view suggests hard pegs, float or float with interventions. 

While conclusions transferred from monetary policy studies should be treated with caution, the bipolar view still offers insights useful for cryptocurrency systems:
Hard pegs using full direct collateralization and floating exchange rate arrangements are less vulnerable to speculative attacks then soft pegs.
This explicitly holds for all soft peg implementations that do not allow for the retraction of most of the money stock in any kind of market situation.

As discussed, currency interventions and open market operations (OMO) that contract money supply by selling limited reserves are definitely concerned.
Self-tokenizing OMO and interest rates on deposits buy back coins against self-issued securities with potentially unlimited supply.
As discussed in Sec.~\ref{subsec:subsec:stab_techniques_focalpoints}, buyers of these special-purpose tokens are incentivized by a share in newly minted money in the case of \emph{longterm} increasing demand for the stablecoin.
As discussed in Sec.~\ref{subsec:era_speculativeattacks} speculative attacks entail almost no risk or costs for the attacker, making repeated attempts in the \emph{short run} possible.
While in the presence of speculators for and against the peg the first series of speculative attacks might be neutralized, claims for risk remuneration will stack up quickly.
Leading to a decrease in relative ownership of future remuneration, this will decrease the demand for the used special-purpose tokens with every round of attack.
Missing demand for the self-issued tokens makes it impossible to absorb money supply and defend against the attack.
Further research is strongly encouraged as the above setup is quite popular.
Nine out of the 24 reviewed projects and thus almost \SI{38}{\percent} consider it.

We do question though, if all kinds of soft pegs are equally vulnerable in the case of cryptocurrencies.
Soft-pegs relying solely on full proxy and self-collateralization promise to provide sufficient collateral to buy back the complete stock of money at any moment of time.
This, in turn, makes them immune to the above described attack~\cite{obstfeld1995mirage}.

\section{Monetary regimes}
\label{sec:mr}
Up to this point, we used the notion of ``stability'' in the sense of low exchange rate volatility.
In the following, we zoom out further, stressing the difference between
\begin{itemize}
\item stabilizing the \emph{amount of another currency} one cryptocurrency unit can buy (exchange rate) and
\item stabilizing the \emph{amount of goods and services} one cryptocurrency unit can buy (purchasing power).
\end{itemize}

Stable purchasing power is a goal which traditional central banks and stablecoins both pursue.
Stability of prices can be measured, \eg, through a basket of goods in a consumer price index (CPI).
In practice it is can be influenced only via indirect measures.
These encompass interest rates, exchange rates and many others.
The respective choice of tool set constitutes the monetary regime. 
Each monetary regime chooses a certain core variable, the so-called \emph{nominal anchor}, to construct its monetary policy around. 
The chosen nominal anchor is used to choose practical applications of monetary instruments and to evaluate their effectiveness.
It can be seen as the central element of the monetary regime and as the measurement variable around which central bank communication and also accountability line up.

Thus, while stable purchasing power is the overarching goal---fixing exchange rates (so called \emph{exchange rate targeting}) is only one of several strategies to achieve it.
Other monetary regimes focus on other factors than the exchange rate, namely \emph{monetary targeting} and \emph{inflation targeting}. 

\emph{Monetary targeting} uses the amount of money as its nominal anchor \cite{miskin2007monetary}.
Assuming predictable velocity of money, the so-called Quantity Equation of Money can be used to calculate the necessary money supply to achieve a certain level of prices~\cite{friedman2017quantity}.\footnote{
  Different versions of the quantity equation of money arose after being popularized by~\cite{fisher1911equation}.
  All have in common that they relate the aggregated flows of money to aggregated flows of goods and services.
  The equations offer different perspectives on the demand of money.
}
Correspondingly, adjusting the money supply is a key means of intervention for a central bank in such regimes.

\emph{Inflation targeting} uses the change in a consumer price index as nominal anchor \cite{bernanke1999inflation}.
The most characteristic differences to monetary targeting lies in the publication of numerical inflation targets and the commitment to hit them.
Additionally, also commitment and ability to achieve the inflation target, emphasis on transparency and increased accountability are quoted characteristic of inflation targeting~\cite{miskin2007monetary}.

\subsection{Regime-inherent aspects}
\label{subsec:mt_ert}
While \emph{exchange rate targeting} is a popular arrangement for cryptocurrencies and countries alike, it exhibits major drawbacks.
First, as stated in~\cite{obstfeld1995mirage}, exchange rate targeters lose the ability to pursue independent monetary policy.
Moreover, inflationary tendencies and shocks are imported directly into the cryptocurrency.
Third, as discussed in Sec.~\ref{subsec:peg_hard}, exchange rate targeting can lead to vulnerabilities to speculative attacks.
On the other hand, exchange rate targeting offers convincing advantages from the perspective of cryptocurrencies.
First, pegging the value of a cryptocurrency to some other currency or asset can reduce the volatility drastically, since price fluctuations of, \eg, USD, are magnitudes smaller than in most cryptocurrencies~\cite{yermack2015bitcoin}.
Second, not even the most mature cryptocurrencies do succeed to be used as unit of account for the purchase of goods or services~\cite{yermack2015bitcoin,glaser2014bitcoin}, so that prices are not typically quoted in cryptocurrency units.
Therefore, also from a usability perspective it is a reasonable choice to strive for a stable relationship to fiat currencies.

\emph{Inflation targeting} poses the obvious challenges of the definition and tracking of an adequate basket of goods and services.
More importantly, goods usually are denominated in some national currency.
Purchasing power fluctuations of a volatile cryptocurrency measured by a basket of dollar denominated goods should mainly be caused by exchange rate variability.
As a consequence inflation targeting and exchange rate targeting would be close to equivalent.

\emph{Monetary targeting} in a sense is already implemented by traditional cryptocurrencies with predetermined block reward.
More sophisticated monetary targeting approaches might use literature around rule based monetary policy (\eg~\cite{mccallum1984credibility}, \cite{mccallum1999issues}, \cite{levin1999robustness} or \cite{mccallum1988robustness}) as first starting point.
If additional measures against exchange rates are to be taken, managed floating regimes as promoted by~\cite{larrain2002should,goldstein2002managed}, might offer a simple but sustainable alternative to exchange rate targeting.
This approach, however, mitigates only short term fluctuations. 

\subsection{Classification results and blank spots}
\label{subsec:mt_blankspots}
\begin{figure}
\begin{tikzpicture}[x=1pt,y=1pt]
\definecolor{fillColor}{RGB}{255,255,255}
\path[use as bounding box,fill=fillColor,fill opacity=0.00] (0,0) rectangle (245.72,180.67);
\begin{scope}
\path[clip] (  0.00,  0.00) rectangle (245.72,180.67);
\definecolor{drawColor}{RGB}{255,255,255}
\definecolor{fillColor}{RGB}{255,255,255}

\path[draw=drawColor,line width= 0.5pt,line join=round,line cap=round,fill=fillColor] ( -0.00,  0.00) rectangle (245.72,180.68);
\end{scope}
\begin{scope}
\path[clip] ( 71.26, 62.74) rectangle (240.72,175.67);
\definecolor{fillColor}{RGB}{255,255,255}

\path[fill=fillColor] ( 71.26, 62.74) rectangle (240.72,175.67);
\definecolor{drawColor}{gray}{0.92}

\path[draw=drawColor,line width= 0.3pt,line join=round] ( 98.22, 62.74) --
  ( 98.22,175.67);

\path[draw=drawColor,line width= 0.3pt,line join=round] (136.73, 62.74) --
  (136.73,175.67);

\path[draw=drawColor,line width= 0.3pt,line join=round] (175.25, 62.74) --
  (175.25,175.67);

\path[draw=drawColor,line width= 0.3pt,line join=round] (213.76, 62.74) --
  (213.76,175.67);

\path[draw=drawColor,line width= 0.5pt,line join=round] ( 71.26, 78.87) --
  (240.72, 78.87);

\path[draw=drawColor,line width= 0.5pt,line join=round] ( 71.26,105.76) --
  (240.72,105.76);

\path[draw=drawColor,line width= 0.5pt,line join=round] ( 71.26,132.65) --
  (240.72,132.65);

\path[draw=drawColor,line width= 0.5pt,line join=round] ( 71.26,159.54) --
  (240.72,159.54);

\path[draw=drawColor,line width= 0.5pt,line join=round] ( 78.97, 62.74) --
  ( 78.97,175.67);

\path[draw=drawColor,line width= 0.5pt,line join=round] (117.48, 62.74) --
  (117.48,175.67);

\path[draw=drawColor,line width= 0.5pt,line join=round] (155.99, 62.74) --
  (155.99,175.67);

\path[draw=drawColor,line width= 0.5pt,line join=round] (194.50, 62.74) --
  (194.50,175.67);

\path[draw=drawColor,line width= 0.5pt,line join=round] (233.02, 62.74) --
  (233.02,175.67);
\definecolor{fillColor}{RGB}{166,97,26}

\path[fill=fillColor] ( 78.97, 82.91) rectangle (111.06, 90.97);
\definecolor{fillColor}{RGB}{223,194,125}

\path[fill=fillColor] ( 78.97, 74.84) rectangle ( 78.97, 82.91);
\definecolor{fillColor}{RGB}{1,133,113}

\path[fill=fillColor] ( 78.97, 66.77) rectangle ( 78.97, 74.84);
\definecolor{fillColor}{RGB}{166,97,26}

\path[fill=fillColor] ( 78.97,109.80) rectangle (115.87,117.86);
\definecolor{fillColor}{RGB}{223,194,125}

\path[fill=fillColor] ( 78.97,101.73) rectangle ( 78.97,109.80);
\definecolor{fillColor}{RGB}{1,133,113}

\path[fill=fillColor] ( 78.97, 93.66) rectangle ( 78.97,101.73);
\definecolor{fillColor}{RGB}{166,97,26}

\path[fill=fillColor] ( 78.97,136.69) rectangle ( 98.22,144.75);
\definecolor{fillColor}{RGB}{223,194,125}

\path[fill=fillColor] ( 78.97,128.62) rectangle ( 85.38,136.69);
\definecolor{fillColor}{RGB}{1,133,113}

\path[fill=fillColor] ( 78.97,120.55) rectangle ( 91.80,128.62);
\definecolor{fillColor}{RGB}{166,97,26}

\path[fill=fillColor] ( 78.97,163.57) rectangle (144.76,171.64);
\definecolor{fillColor}{RGB}{223,194,125}

\path[fill=fillColor] ( 78.97,155.51) rectangle (168.83,163.57);
\definecolor{fillColor}{RGB}{1,133,113}

\path[fill=fillColor] ( 78.97,147.44) rectangle (220.18,155.51);
\definecolor{drawColor}{RGB}{0,0,0}

\node[text=drawColor,anchor=base west,inner sep=0pt, outer sep=0pt, scale=  0.85] at (111.82, 83.10) {20.8};

\node[text=drawColor,anchor=base west,inner sep=0pt, outer sep=0pt, scale=  0.85] at ( 79.18, 75.93) {0};

\node[text=drawColor,anchor=base west,inner sep=0pt, outer sep=0pt, scale=  0.85] at ( 79.18, 68.76) {0};

\node[text=drawColor,anchor=base west,inner sep=0pt, outer sep=0pt, scale=  0.85] at (116.30,109.99) {24};

\node[text=drawColor,anchor=base west,inner sep=0pt, outer sep=0pt, scale=  0.85] at ( 79.18,102.82) {0};

\node[text=drawColor,anchor=base west,inner sep=0pt, outer sep=0pt, scale=  0.85] at ( 79.18, 95.65) {0};

\node[text=drawColor,anchor=base west,inner sep=0pt, outer sep=0pt, scale=  0.85] at ( 98.98,136.88) {12.5};

\node[text=drawColor,anchor=base west,inner sep=0pt, outer sep=0pt, scale=  0.85] at ( 85.93,129.71) {4.2};

\node[text=drawColor,anchor=base west,inner sep=0pt, outer sep=0pt, scale=  0.85] at ( 92.35,122.54) {8.3};

\node[text=drawColor,anchor=base west,inner sep=0pt, outer sep=0pt, scale=  0.85] at (145.52,163.77) {42.7};

\node[text=drawColor,anchor=base west,inner sep=0pt, outer sep=0pt, scale=  0.85] at (169.59,156.60) {58.3};

\node[text=drawColor,anchor=base west,inner sep=0pt, outer sep=0pt, scale=  0.85] at (220.94,149.43) {91.7};
\definecolor{drawColor}{gray}{0.20}

\path[draw=drawColor,line width= 0.5pt,line join=round,line cap=round] ( 71.26, 62.74) rectangle (240.72,175.67);
\end{scope}
\begin{scope}
\path[clip] (  0.00,  0.00) rectangle (245.72,180.67);
\definecolor{drawColor}{gray}{0.30}

\node[text=drawColor,anchor=base east,inner sep=0pt, outer sep=0pt, scale=  0.80] at ( 66.76, 76.12) {Inflation T.};

\node[text=drawColor,anchor=base east,inner sep=0pt, outer sep=0pt, scale=  0.80] at ( 66.76,103.01) {Other};

\node[text=drawColor,anchor=base east,inner sep=0pt, outer sep=0pt, scale=  0.80] at ( 66.76,129.90) {Monetary T.};

\node[text=drawColor,anchor=base east,inner sep=0pt, outer sep=0pt, scale=  0.80] at ( 66.76,156.79) {Exch. Rate T.};
\end{scope}
\begin{scope}
\path[clip] (  0.00,  0.00) rectangle (245.72,180.67);
\definecolor{drawColor}{gray}{0.20}

\path[draw=drawColor,line width= 0.5pt,line join=round] ( 68.76, 78.87) --
  ( 71.26, 78.87);

\path[draw=drawColor,line width= 0.5pt,line join=round] ( 68.76,105.76) --
  ( 71.26,105.76);

\path[draw=drawColor,line width= 0.5pt,line join=round] ( 68.76,132.65) --
  ( 71.26,132.65);

\path[draw=drawColor,line width= 0.5pt,line join=round] ( 68.76,159.54) --
  ( 71.26,159.54);
\end{scope}
\begin{scope}
\path[clip] (  0.00,  0.00) rectangle (245.72,180.67);
\definecolor{drawColor}{gray}{0.20}

\path[draw=drawColor,line width= 0.5pt,line join=round] ( 78.97, 60.24) --
  ( 78.97, 62.74);

\path[draw=drawColor,line width= 0.5pt,line join=round] (117.48, 60.24) --
  (117.48, 62.74);

\path[draw=drawColor,line width= 0.5pt,line join=round] (155.99, 60.24) --
  (155.99, 62.74);

\path[draw=drawColor,line width= 0.5pt,line join=round] (194.50, 60.24) --
  (194.50, 62.74);

\path[draw=drawColor,line width= 0.5pt,line join=round] (233.02, 60.24) --
  (233.02, 62.74);
\end{scope}
\begin{scope}
\path[clip] (  0.00,  0.00) rectangle (245.72,180.67);
\definecolor{drawColor}{gray}{0.30}

\node[text=drawColor,anchor=base,inner sep=0pt, outer sep=0pt, scale=  0.80] at ( 78.97, 52.73) {0};

\node[text=drawColor,anchor=base,inner sep=0pt, outer sep=0pt, scale=  0.80] at (117.48, 52.73) {25};

\node[text=drawColor,anchor=base,inner sep=0pt, outer sep=0pt, scale=  0.80] at (155.99, 52.73) {50};

\node[text=drawColor,anchor=base,inner sep=0pt, outer sep=0pt, scale=  0.80] at (194.50, 52.73) {75};

\node[text=drawColor,anchor=base,inner sep=0pt, outer sep=0pt, scale=  0.80] at (233.02, 52.73) {100};
\end{scope}
\begin{scope}
\path[clip] (  0.00,  0.00) rectangle (245.72,180.67);
\definecolor{drawColor}{RGB}{0,0,0}

\node[text=drawColor,anchor=base,inner sep=0pt, outer sep=0pt, scale=  1.00] at (155.99, 41.40) {Fraction of projects and countries in \%};
\end{scope}
\begin{scope}
\path[clip] (  0.00,  0.00) rectangle (245.72,180.67);
\definecolor{drawColor}{RGB}{0,0,0}

\node[text=drawColor,rotate= 90.00,anchor=base,inner sep=0pt, outer sep=0pt, scale=  1.00] at ( 11.89,119.21) {Monetary Regime};
\end{scope}
\begin{scope}
\path[clip] (  0.00,  0.00) rectangle (245.72,180.67);
\definecolor{fillColor}{RGB}{255,255,255}

\path[fill=fillColor] ( 60.11,  5.00) rectangle (251.88, 29.45);
\end{scope}
\begin{scope}
\path[clip] (  0.00,  0.00) rectangle (245.72,180.67);
\definecolor{drawColor}{RGB}{0,0,0}

\node[text=drawColor,anchor=base west,inner sep=0pt, outer sep=0pt, scale=  1.00] at ( 65.11, 13.78) {Type};
\end{scope}
\begin{scope}
\path[clip] (  0.00,  0.00) rectangle (245.72,180.67);
\definecolor{fillColor}{RGB}{255,255,255}

\path[fill=fillColor] ( 92.60, 10.00) rectangle (107.05, 24.45);
\end{scope}
\begin{scope}
\path[clip] (  0.00,  0.00) rectangle (245.72,180.67);
\definecolor{fillColor}{RGB}{1,133,113}

\path[fill=fillColor] ( 93.31, 10.71) rectangle (106.34, 23.74);
\end{scope}
\begin{scope}
\path[clip] (  0.00,  0.00) rectangle (245.72,180.67);
\definecolor{fillColor}{RGB}{255,255,255}

\path[fill=fillColor] (147.85, 10.00) rectangle (162.30, 24.45);
\end{scope}
\begin{scope}
\path[clip] (  0.00,  0.00) rectangle (245.72,180.67);
\definecolor{fillColor}{RGB}{223,194,125}

\path[fill=fillColor] (148.56, 10.71) rectangle (161.59, 23.74);
\end{scope}
\begin{scope}
\path[clip] (  0.00,  0.00) rectangle (245.72,180.67);
\definecolor{fillColor}{RGB}{255,255,255}

\path[fill=fillColor] (211.98, 10.00) rectangle (226.43, 24.45);
\end{scope}
\begin{scope}
\path[clip] (  0.00,  0.00) rectangle (245.72,180.67);
\definecolor{fillColor}{RGB}{166,97,26}

\path[fill=fillColor] (212.69, 10.71) rectangle (225.72, 23.74);
\end{scope}
\begin{scope}
\path[clip] (  0.00,  0.00) rectangle (245.72,180.67);
\definecolor{drawColor}{RGB}{0,0,0}

\node[text=drawColor,anchor=base west,inner sep=0pt, outer sep=0pt, scale=  0.80] at (112.05, 14.47) {All Proj.};
\end{scope}
\begin{scope}
\path[clip] (  0.00,  0.00) rectangle (245.72,180.67);
\definecolor{drawColor}{RGB}{0,0,0}

\node[text=drawColor,anchor=base west,inner sep=0pt, outer sep=0pt, scale=  0.80] at (167.30, 14.47) {Impl. Proj.};
\end{scope}
\begin{scope}
\path[clip] (  0.00,  0.00) rectangle (245.72,180.67);
\definecolor{drawColor}{RGB}{0,0,0}

\node[text=drawColor,anchor=base west,inner sep=0pt, outer sep=0pt, scale=  0.80] at (231.43, 14.47) {IMF};
\end{scope}
\end{tikzpicture}

  \caption{Monetary regimes: reviewed projects and central banks.}
  \label{fig:imf_crypto_comparison}
\end{figure}
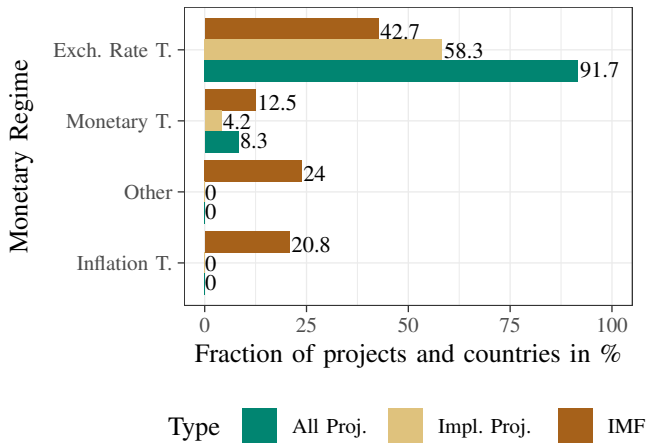

Fig.~\ref{fig:imf_crypto_comparison} shows that the prevalence of exchange rate targeting in the reviewed projects is in stark contrast to traditional central banks
(numbers for countries from \cite{international2016exchange}).
Exchange rate targeting, accounting for \SI{91.7}{\percent} of all projects, is the clear favorite of current approaches to cryptocurrency stabilization.
Only \SI{42.7}{\percent} of countries use a certain exchange rate as their currency's nominal anchor.
While monetary targeting in combination with short term exchange rate smoothing could be an interesting alternative, it has largely been ignored by cryptocurrencies.

\section{Decentralization and Trust}%
\label{sec:dec_trust}

Departing from mainly economic questions, we will now discuss the design of stable cryptocurrencies in terms of their "decentralization" and "trustlessness"---notions that cryptocurrencies are commonly associated with and potentially owe their popularity to.
We focus on contrasting fully \emph{permissionless} systems, \ie, systems that function without previous assumptions about the identifiability of participants or their trustworthiness, with architectures in which the existence of a group of well-known \emph{trusted actors} must be assumed.
With the current state of knowledge, it is an outstanding question whether an effectively price-stable cryptocurrency can at all be realized in a fully permissionless setting.

Techniques based on the collateralization or holding of "off-chain" assets (such as classical currencies) are \emph{inherently incompatible} with a fully permissionless setup.
Neither the actual existence nor the correct management of off-chain collateral can, in general, be secured through purely technical means.
Rather than that, a form of trust, either in one ore more well-known custodians or in a surrounding legal framework and its enforcement mechanics, must be assumed.

Techniques based on collateralization, interest rates and OMOs are in principle compatible with a fully permissionless mode of operation as long as they act on assets whose ownership can be securely tracked and managed in a permissionless manner (\eg, are recorded on the same permissionless ledger).
Even then, however, a fully permissionless mode of operation is only possible under a significant caveat---the existence of a secure (permissionless) \emph{oracle} for the chosen nominal anchor.
Oracles are system components that transfer "external" information onto the blockchain.
Monetary information, such as the price of the cryptocurrency relative to another currency, are required for monetary policy mechanisms.
They are typically not natively generated "on-chain" and must therefore be provided by an oracle.
Oracles can be trivially realized using a trusted party that vouches for the correctness of data by means of cryptographic signatures.
However, this clearly reinstates a globally trusted actor (or a group thereof).
Completely permissionless oracles are, on the other hand, still an active research field, with no sufficiently secure solutions for, \eg, transferring price data, currently in sight~\cite{DBLP:journals/corr/abs-1808-00528}.
It is possible that secure permissionless oracles for arbitrary data are a theoretical impossibility~\cite{douceur2002sybil}.
In such a case, a final possibility for the realization of completely permissionless stablecoins remains in the deepened investigation of on-chain \emph{proxy variables} for relevant nominal anchors like current prices.
We are currently aware only of the current mining hash rate, as materialized, \eg, in the timing between blocks and the current mining difficulty, as a potentially viable representative of this class.
More research is needed here to further test the viability of this approach, especially in respect to incentive-compatibility, gameability and security (c.f. Sec.~\ref{subsec:subsec:stab_techniques_focalpoints:mining}).

\section{From individual classifications to taxonomy}
\label{sec:mt_jointaxonomy}
In the course of this article, we put together three perspectives on price stability: (i) practical stabilization techniques, (ii) exchange rate regimes, (iii) monetary regimes.
Each perspective offers a classification.
Taken together, the three classifications build a taxonomy that might prove a helpful tool for developers, research and investors.
A schematic overview over can be found in Appendix \ref{appendix:taxonomy}

\section{Conclusion}
\label{sec:conclusion}
In this paper, we systematically explore the enigma of monetary stabilization in cryptocurrencies.
We go beyond individual proposals, focusing on overarching concepts and approaches.
We extracted information from 24 stablecoin projects and combine the resulting insights with economic literature yielding a comprehensive taxonomy for the analysis and classification of stablecoins.
We find that, the three most popular stability techniques following after direct collateralization are unsustainable in the long run.
Moreover, our findings show that almost \SI{38}{\percent} of surveyed coins promote a problematic combination of exchange rate targeting and using either limited reserves or a potentially unlimited supply of self-issued tokens to reduce the coin supply.
There are strong indicators that the above setup can result in a vulnerability to speculative attacks.
On the other hand, proxy and self collateralization (promising alternative techniques that might be applicable in ``trustless'' settings) rely heavily on margin calls with questionable robustness.
Further research is required to evaluate the viability of such margin calls in small and potentially illiquid markets.
Zooming out, we suggest that short term smoothing of exchange rates might offer a sustainable alternative to exchange rate targeting---the current focus of over \SI{90}{\percent} of projects.

We identify a number of further opportunities for technical and economics research on cryptocurrency stabilization, such as on the resilience of self-tokenizing techniques, on the viability of secure permissionless price feeds for informing policing decisions, and on the actual effectiveness of monetary policy given the complete transparency of both the policy and its enforcement.

\clearpage

\onecolumn
\appendices

\section{Surveyed projects}
\label{appendix:surveyed_projects}
\begin{flushleft}
  A list of surveyed stablecoin projects and their respective classification. Note that, ``partially implemented'' refers to the fact that the coin itself is traded while not all announced stabilization techniques are implemented yet.
\end{flushleft}
  
\begin{centering} 
  \begin{adjustbox}{
      width=\textwidth,
      totalheight=\textheight,
      keepaspectratio,
      rotate=0,
      nofloat=table}
    \begin{tabular}{@{\extracolsep{5pt}} ccccccccccccccc} 
\\[-1.8ex]\hline 
\hline \\[-1.8ex] 
Project (Stabilized Token) & Status & MR & ERA & T1 & T2 & T3 & T4 & T5 & T6 & T7 & T8 & T9 & T10 & T11 \\ 
\hline \\[-1.8ex] 
Augmint (A-EUR) & not impl. & ERT & soft peg & - & - & - & - & yes & yes & - & yes & - & - & - \\ 
Aurora (Boreal) & not impl. & ERT & soft peg & - & - & - & - & yes & - & - & yes & - & - & - \\ 
Basecoin (Basis) & retracted & ERT & soft peg & - & - & - & - & - & - & - & - & yes & yes & - \\ 
BitShares (BitUSD) & impl. & ERT & soft peg & - & - & yes & - & - & - & - & - & - & - & - \\ 
Carbon (Carbon) & impl. & ERT & soft peg & - & - & - & - & - & - & - & - & yes & yes & - \\ 
Celo (Celo) & not impl. & ERT & soft peg & - & - & - & - & - & - & - & yes & yes & yes & - \\ 
Centre (USD Coin) & impl. & ERT & hard peg & yes & - & - & - & - & - & - & - & - & - & - \\ 
Digix (Digix Gold Token) & impl. & ERT & hard peg & yes & - & - & - & - & - & - & - & - & - & - \\ 
Fragments (Fragments) & not impl. & ERT & soft peg & - & - & - & - & - & - & - & yes & yes & - & - \\ 
Globcoin (GLC Token) & impl. & ERT & hard peg & yes & - & - & - & - & - & - & - & - & - & - \\ 
Karbo (Karbo) & partially impl. & MT & free float & - & - & - & - & - & yes & - & - & - & yes & - \\ 
Kowala (kUSD) & not impl. & ERT & soft peg & - & - & - & - & - & - & - & - & - & yes & yes \\ 
Maker (Dai) & impl. & ERT & soft peg & - & - & - & - & yes & - & - & - & - & - & - \\ 
Minex Coin (Minex Coin) & impl. & MT & float. w. int. & - & - & - & yes & - & yes & - & - & - & - & - \\ 
NuBits (Nubits) & impl. & ERT & soft peg & - & - & - & yes & - & yes & - & - & - & yes & - \\ 
Stableunit (Stableunit) & not impl. & ERT & soft peg & - & - & - & - & - & yes & - & yes & yes & - & - \\ 
Stably (StableUSD) & impl. & ERT & hard peg & yes & - & - & - & - & - & - & - & - & - & - \\ 
Stasis (EURS) & impl. & ERT & hard peg & yes & - & - & - & - & - & - & - & - & - & - \\ 
Stronghold (USDS) & impl. & ERT & hard peg & yes & - & - & - & - & - & - & - & - & - & - \\ 
Sythetix (SUSD) & impl. & ERT & soft peg & - & - & yes & - & - & - & - & - & - & - & - \\ 
Tether (USDT) & impl. & ERT & hard peg & yes & - & - & - & - & - & - & - & - & - & - \\ 
Trusttoken (TrueUSD) & impl. & ERT & hard peg & yes & - & - & - & - & - & - & - & - & - & - \\ 
USC (USC) & impl. & ERT & hard peg & yes & - & - & - & - & - & - & - & - & - & - \\ 
x8currency (X8C) & not impl. & ERT & hard peg & yes & - & - & - & - & - & - & - & - & - & - \\ 
\hline \\[-1.8ex] 
\end{tabular}
  \end{adjustbox}
  \begin{adjustbox}{
      width=\textwidth,
      totalheight=\textheight,
      keepaspectratio,
      rotate=0,
      nofloat=table}
    \begin{tabular*}{\textwidth}{ll}
      & \\
      \textbf{Abbreviation} & \textbf{Full text} \\
      MR & Monetary regime\\
      ERA & Exchange range arrangement\\
      MT & Monetary targeting\\
      ERT & Exchange rate targeting\\
      float. w. int. & Floating with interventions\\
      impl. & Implemented\\
      T1 & Collateralization (direct)\\
      T2 & Collateralization (proxy)\\
      T3 & Collateralization (self)\\
      T4 & Currency interventions\\
      T5 & Interest rates with loans\\
      T6 & Interest rates with deposits\\
      T7 & Open market operations (standard)\\
      T8 & Open market operations (proxy)\\
      T9 & Open market operations (self-tokenizing)\\
      T10 & Dynamic mining reward\\
      T11 & Dynamically burned transaction fee\\
    \end{tabular*}  
  \end{adjustbox}
\end{centering}

\clearpage

\section{Taxonomy}
\label{appendix:taxonomy}

\begin{figure*}[ht!]
  \centering
  \usetikzlibrary{calc}
\tikzset{
  every point/.style = {radius={\pgflinewidth}, opacity=1, draw, solid, fill=white},
  pt/.pic = {
    \begin{pgfonlayer}{foreground}
      \path[every point, #1] circle;
    \end{pgfonlayer}
  },
  point/.style={insert path={pic{pt={#1}}}}, point/.default={},
  point name/.style = {insert path={coordinate (#1)}}
}

\pgfdeclarelayer{background}%
\pgfdeclarelayer{foreground}%
\pgfsetlayers{background,main,foreground}%

\begin{tikzpicture}[]
  \def \debugPoint   {}%
  \def \colorOut     {black!100!white}
  \def \colorInA     {black!60!white}
  \def \colorInB     {black!40!white}
  \def \lineOutW     {1.0pt}
  \def \lineInW      {0.7pt}
  \def \lineInAW     {1.0pt}
  \def \lineInBW     {0.7pt}
  \def \picW         {\linewidth*29.5/31}%
  \def \picH         {\paperheight*13.75/31}%
  \def \bOutDistW    {0.2cm}
  \def \bOutDistH    {\bOutDistW}
  
  \def \bOutSmallW   {\picW*0.11}
  \def \bOutSmallWH  {\picW*0.0505}
  \def \bOutSmallH   {\picH*0.3}
  \def \bOutSmallHH  {\picH*0.15}
  
  \def \bOutBigW     {\picW*0.78}
  \def \bOutBigWH    {\picW*0.4}
  \def \bOutBigH     {\bOutSmallH}
  \def \bOutBigHH    {\bOutSmallHH}
  
  \def \bInDistAW    {0.28cm}
  \def \bInDistAH    {\bInDistAW}
  \def \bInDistBW    {0.1cm}
  \def \bInDistBH    {\bInDistBW}
  
  \def \bInBigWV     {\bOutBigW*1/3-\bInDistAW*2/3-\bInDistBW*2/3}
  \def \bInBigWH     {\bOutBigW*1/6-\bInDistAW*2/6-\bInDistBW*2/6}
  \def \bInBigHV     {\bOutBigH*1/1-\bInDistAH*2/1}
  \def \bInBigHH     {\bOutBigH*1/2-\bInDistAH*2/2}

  \def \bInSmallWV   {\bOutBigW*1/7-\bInDistAW*2/7-\bInDistBW*6/7}
  \def \bInSmallWH   {\bOutBigW*1/14-\bInDistAW*2/14-\bInDistBW*6/14}
  \def \bInSmallHV   {\bOutBigH*1/2-\bInDistAH*2/2-\bInDistBW*1/2}
  \def \bInSmallHH   {\bOutBigH*1/4-\bInDistAH*2/4-\bInDistBW*1/4}
  
  \def \bInSmallBAWV {\bOutBigW*1/12-\bInDistAW*2/12-\bInDistBW*2/12-\bInDistBW*1/4}
  \def \bInSmallBAWH {\bOutBigW*1/24-\bInDistAW*2/24-\bInDistBW*2/24-\bInDistBW*1/8}
  \def \bInSmallBAHV {\bOutBigH*1/3-\bInDistAH*2/3-\bInDistBW*2/3}
  \def \bInSmallBAHH {\bOutBigH*1/6-\bInDistAH*2/6-\bInDistBW*2/6}
  
  \def \bInSmallBBWV {\bOutBigW*3/12-\bInDistAW*6/12-\bInDistBW*6/12-\bInDistBW*3/4}
  \def \bInSmallBBWH {\bOutBigW*3/24-\bInDistAW*6/24-\bInDistBW*6/24-\bInDistBW*3/8}
  \def \bInSmallBBHV {\bOutBigH*1/2-\bInDistAH*2/2-\bInDistBW*1/2}
  \def \bInSmallBBHH {\bOutBigH*1/4-\bInDistAH*2/4-\bInDistBW*1/4}
  \coordinate (CBOAA)     at ($(0,0)     + ( 0.0             , 0.0             )$) (CBOAA)   [\debugPoint];%
  \coordinate (CBOAB)     at ($(CBOAA)   + ( 0.0             , \bOutSmallH     )$) (CBOAB)   [\debugPoint];%
  \coordinate (CBOAC)     at ($(CBOAB)   + ( 0.0             , \bOutDistH      )$) (CBOAC)   [\debugPoint];%
  \coordinate (CBOAD)     at ($(CBOAC)   + ( 0.0             , \bOutSmallH     )$) (CBOAD)   [\debugPoint];%
  \coordinate (CBOAE)     at ($(CBOAD)   + ( 0.0             , \bOutDistH      )$) (CBOAE)   [\debugPoint];%
  \coordinate (CBOAF)     at ($(CBOAE)   + ( 0.0             , \bOutSmallH     )$) (CBOAF)   [\debugPoint];%
  
  \coordinate (CBOBA)     at ($(CBOAA)   + ( \bOutSmallW     , 0.0             )$) (CBOBA)   [\debugPoint];%
  \coordinate (CBOBB)     at ($(CBOBA)   + ( 0.0             , \bOutSmallH     )$) (CBOBB)   [\debugPoint];%
  \coordinate (CBOBC)     at ($(CBOBB)   + ( 0.0             , \bOutDistH      )$) (CBOBC)   [\debugPoint];%
  \coordinate (CBOBD)     at ($(CBOBC)   + ( 0.0             , \bOutSmallH     )$) (CBOBD)   [\debugPoint];%
  \coordinate (CBOBE)     at ($(CBOBD)   + ( 0.0             , \bOutDistH      )$) (CBOBE)   [\debugPoint];%
  \coordinate (CBOBF)     at ($(CBOBE)   + ( 0.0             , \bOutSmallH     )$) (CBOBF)   [\debugPoint];%
  
  \coordinate (CBOCA)     at ($(CBOBA)   + ( \bOutDistW      , 0.0             )$) (CBOCA)   [\debugPoint];%
  \coordinate (CBOCB)     at ($(CBOCA)   + ( 0.0             , \bOutSmallH     )$) (CBOCB)   [\debugPoint];%
  \coordinate (CBOCC)     at ($(CBOCB)   + ( 0.0             , \bOutDistH      )$) (CBOCC)   [\debugPoint];%
  \coordinate (CBOCD)     at ($(CBOCC)   + ( 0.0             , \bOutSmallH     )$) (CBOCD)   [\debugPoint];%
  \coordinate (CBOCE)     at ($(CBOCD)   + ( 0.0             , \bOutDistH      )$) (CBOCE)   [\debugPoint];%
  \coordinate (CBOCF)     at ($(CBOCE)   + ( 0.0             , \bOutSmallH     )$) (CBOCF)   [\debugPoint];%
  
  \coordinate (CBODA)     at ($(CBOCA)   + ( \bOutSmallW     , 0.0             )$) (CBODA)   [\debugPoint];%
  \coordinate (CBODB)     at ($(CBODA)   + ( 0.0             , \bOutSmallH     )$) (CBODB)   [\debugPoint];%
  \coordinate (CBODC)     at ($(CBODB)   + ( 0.0             , \bOutDistH      )$) (CBODC)   [\debugPoint];%
  \coordinate (CBODD)     at ($(CBODC)   + ( 0.0             , \bOutSmallH     )$) (CBODD)   [\debugPoint];%
  \coordinate (CBODE)     at ($(CBODD)   + ( 0.0             , \bOutDistH      )$) (CBODE)   [\debugPoint];%
  \coordinate (CBODF)     at ($(CBODE)   + ( 0.0             , \bOutSmallH     )$) (CBODF)   [\debugPoint];%
  
  \coordinate (CBOEA)     at ($(CBODA)   + ( \bOutDistW      , 0.0             )$) (CBOEA)   [\debugPoint];%
  \coordinate (CBOEB)     at ($(CBOEA)   + ( 0.0             , \bOutSmallH     )$) (CBOEB)   [\debugPoint];%
  \coordinate (CBOEC)     at ($(CBOEB)   + ( 0.0             , \bOutDistH      )$) (CBOEC)   [\debugPoint];%
  \coordinate (CBOED)     at ($(CBOEC)   + ( 0.0             , \bOutSmallH     )$) (CBOED)   [\debugPoint];%
  \coordinate (CBOEE)     at ($(CBOED)   + ( 0.0             , \bOutDistH      )$) (CBOEE)   [\debugPoint];%
  \coordinate (CBOEF)     at ($(CBOEE)   + ( 0.0             , \bOutSmallH     )$) (CBOEF)   [\debugPoint];%
  
  \coordinate (CBOFA)     at ($(CBOEA)   + ( \bOutBigW       , 0.0             )$) (CBOFA)   [\debugPoint];%
  \coordinate (CBOFB)     at ($(CBOFA)   + ( 0.0             , \bOutSmallH     )$) (CBOFB)   [\debugPoint];%
  \coordinate (CBOFC)     at ($(CBOFB)   + ( 0.0             , \bOutDistH      )$) (CBOFC)   [\debugPoint];%
  \coordinate (CBOFD)     at ($(CBOFC)   + ( 0.0             , \bOutSmallH     )$) (CBOFD)   [\debugPoint];%
  \coordinate (CBOFE)     at ($(CBOFD)   + ( 0.0             , \bOutDistH      )$) (CBOFE)   [\debugPoint];%
  \coordinate (CBOFF)     at ($(CBOFE)   + ( 0.0             , \bOutSmallH     )$) (CBOFF)   [\debugPoint];%
  
  \coordinate (CBIAA)     at ($(CBOEA)   + ( \bInDistAH       , \bInDistAH     )$) (CBIAA)   [\debugPoint];%
    \coordinate (CBIAB)     at ($(CBIAA)   + ( 0.0              , \bInSmallHV    )$) (CBIAB)   [\debugPoint];%
    \coordinate (CBIAC)     at ($(CBIAB)   + ( 0.0              , \bInDistBW     )$) (CBIAC)   [\debugPoint];%
  \coordinate (CBIAD)     at ($(CBIAC)   + ( 0.0              , \bInSmallHV    )$) (CBIAD)   [\debugPoint];%
  \coordinate (CBIAE)     at ($(CBOEC)   + ( \bInDistAW       , \bInDistAH     )$) (CBIAE)   [\debugPoint];%
  \coordinate (CBIAF)     at ($(CBIAE)   + ( 0.0              , \bInBigHV      )$) (CBIAF)   [\debugPoint];%
  \coordinate (CBIAG)     at ($(CBOEE)   + ( \bInDistAW       , \bInDistAH     )$) (CBIAG)   [\debugPoint];%
  \coordinate (CBIAH)     at ($(CBIAG)   + ( 0.0              , \bInBigHV      )$) (CBIAH)   [\debugPoint];%
  
  \coordinate (CBIBA)     at ($(CBIAA)   + ( \bInSmallWV      , 0.0            )$) (CBIBA)   [\debugPoint];%
  \coordinate (CBIBB)     at ($(CBIAB)   + ( \bInSmallWV      , 0.0            )$) (CBIBB)   [\debugPoint];%
  \coordinate (CBIBC)     at ($(CBIAC)   + ( \bInSmallWV      , 0.0            )$) (CBIBC)   [\debugPoint];%
  \coordinate (CBIBD)     at ($(CBIAD)   + ( \bInSmallWV      , 0.0            )$) (CBIBD)   [\debugPoint];%
  \coordinate (CBIBE)     at ($(CBIAE)   + ( \bInSmallBAWV    , 0.0            )$) (CBIBE)   [\debugPoint];%
  \coordinate (CBIBF)     at ($(CBIAF)   + ( \bInSmallBAWV    , 0.0            )$) (CBIBF)   [\debugPoint];%
  \coordinate (CBIBG)     at ($(CBIAG)   + ( \bInBigWV        , 0.0            )$) (CBIBG)   [\debugPoint];%
  \coordinate (CBIBH)     at ($(CBIAH)   + ( \bInBigWV        , 0.0            )$) (CBIBH)   [\debugPoint];%
  
  \coordinate (CBICA)     at ($(CBIBA)   + ( \bInDistBW       , 0.0            )$) (CBICA)   [\debugPoint];%
  \coordinate (CBICB)     at ($(CBIBB)   + ( \bInDistBW       , 0.0            )$) (CBICB)   [\debugPoint];%
  \coordinate (CBICC)     at ($(CBIBC)   + ( \bInDistBW       , 0.0            )$) (CBICC)   [\debugPoint];%
  \coordinate (CBICD)     at ($(CBIBD)   + ( \bInDistBW       , 0.0            )$) (CBICD)   [\debugPoint];%
  \coordinate (CBICE)     at ($(CBIBE)   + ( \bInDistBW       , 0.0            )$) (CBICE)   [\debugPoint];%
  \coordinate (CBICEA)    at ($(CBICE)   + ( 0.0              , \bInSmallBBHV  )$) (CBICEA)  [\debugPoint];%
  \coordinate (CBICEB)    at ($(CBICEA)  + ( 0.0              , \bInDistBH     )$) (CBICEB)  [\debugPoint];%
  \coordinate (CBICF)     at ($(CBIBF)   + ( \bInDistBW       , 0.0            )$) (CBICF)   [\debugPoint];%
  \coordinate (CBICG)     at ($(CBIBG)   + ( \bInDistBW       , 0.0            )$) (CBICG)   [\debugPoint];%
  \coordinate (CBICH)     at ($(CBIBH)   + ( \bInDistBW       , 0.0            )$) (CBICH)   [\debugPoint];%
  
  \coordinate (CBIDA)     at ($(CBICA)   + ( \bInSmallWV      , 0.0            )$) (CBIDA)   [\debugPoint];%
  \coordinate (CBIDB)     at ($(CBICB)   + ( \bInSmallWV      , 0.0            )$) (CBIDB)   [\debugPoint];%
  \coordinate (CBIDC)     at ($(CBICC)   + ( \bInSmallWV      , 0.0            )$) (CBIDC)   [\debugPoint];%
  \coordinate (CBIDD)     at ($(CBICD)   + ( \bInSmallWV      , 0.0            )$) (CBIDD)   [\debugPoint];%
  \coordinate (CBIDE)     at ($(CBIAE)   + ( \bInBigWV        , 0.0            )$) (CBIDE)   [\debugPoint];%
  \coordinate (CBIDEA)    at ($(CBIDE)   + ( 0.0              , \bInSmallBBHV  )$) (CBIDEA)  [\debugPoint];%
  \coordinate (CBIDEB)    at ($(CBIDEA)  + ( 0.0              , \bInDistBH     )$) (CBIDEB)  [\debugPoint];%
  \coordinate (CBIDF)     at ($(CBIAF)   + ( \bInBigWV        , 0.0            )$) (CBIDF)   [\debugPoint];%
  \coordinate (CBIDG)     at ($(CBICG)   + ( \bInBigWV        , 0.0            )$) (CBIDG)   [\debugPoint];%
  \coordinate (CBIDH)     at ($(CBICH)   + ( \bInBigWV        , 0.0            )$) (CBIDH)   [\debugPoint];%
  
  \coordinate (CBIEA)     at ($(CBIDA)   + ( \bInDistBW       , 0.0            )$) (CBIEA)   [\debugPoint];%
  \coordinate (CBIEB)     at ($(CBIDB)   + ( \bInDistBW       , 0.0            )$) (CBIEB)   [\debugPoint];%
  \coordinate (CBIEC)     at ($(CBIDC)   + ( \bInDistBW       , 0.0            )$) (CBIEC)   [\debugPoint];%
  \coordinate (CBIED)     at ($(CBIDD)   + ( \bInDistBW       , 0.0            )$) (CBIED)   [\debugPoint];%
  \coordinate (CBIEE)     at ($(CBIDE)   + ( \bInDistBW       , 0.0            )$) (CBIEE)   [\debugPoint];%
  \coordinate (CBIEF)     at ($(CBIDF)   + ( \bInDistBW       , 0.0            )$) (CBIEF)   [\debugPoint];%
  \coordinate (CBIEG)     at ($(CBIDG)   + ( \bInDistBW       , 0.0            )$) (CBIEG)   [\debugPoint];%
  \coordinate (CBIEH)     at ($(CBIDH)   + ( \bInDistBW       , 0.0            )$) (CBIEH)   [\debugPoint];%
  
  \coordinate (CBIFA)     at ($(CBIEA)   + ( \bInSmallWV      , 0.0            )$) (CBIFA)   [\debugPoint];%
  \coordinate (CBIFB)     at ($(CBIEB)   + ( \bInSmallWV      , 0.0            )$) (CBIFB)   [\debugPoint];%
  \coordinate (CBIFC)     at ($(CBIEC)   + ( \bInSmallWV      , 0.0            )$) (CBIFC)   [\debugPoint];%
  \coordinate (CBIFD)     at ($(CBIED)   + ( \bInSmallWV      , 0.0            )$) (CBIFD)   [\debugPoint];%
  \coordinate (CBIFE)     at ($(CBIEE)   + ( \bInSmallBAWV    , 0.0            )$) (CBIFE)   [\debugPoint];%
  \coordinate (CBIFF)     at ($(CBIEF)   + ( \bInSmallBAWV    , 0.0            )$) (CBIFF)   [\debugPoint];%
  \coordinate (CBIFG)     at ($(CBIEG)   + ( \bInBigWV        , 0.0            )$) (CBIFG)   [\debugPoint];%
  \coordinate (CBIFH)     at ($(CBIEH)   + ( \bInBigWV        , 0.0            )$) (CBIFH)   [\debugPoint];%
  
  \coordinate (CBIGA)     at ($(CBIFA)   + ( \bInDistBW       , 0.0            )$) (CBIGA)   [\debugPoint];%
  \coordinate (CBIGB)     at ($(CBIFB)   + ( \bInDistBW       , 0.0            )$) (CBIGB)   [\debugPoint];%
  \coordinate (CBIGC)     at ($(CBIFC)   + ( \bInDistBW       , 0.0            )$) (CBIGC)   [\debugPoint];%
  \coordinate (CBIGD)     at ($(CBIFD)   + ( \bInDistBW       , 0.0            )$) (CBIGD)   [\debugPoint];%
  \coordinate (CBIGE)     at ($(CBIFE)   + ( \bInDistBW       , 0.0            )$) (CBIGE)   [\debugPoint];%
  \coordinate (CBIGEA)    at ($(CBIGE)   + ( 0.0              , \bInSmallBAHV  )$) (CBIGEA)  [\debugPoint];%
  \coordinate (CBIGEB)    at ($(CBIGEA)  + ( 0.0              , \bInDistBH     )$) (CBIGEB)  [\debugPoint];%
  \coordinate (CBIGEC)    at ($(CBIGEB)  + ( 0.0              , \bInSmallBAHV  )$) (CBIGEC)  [\debugPoint];%
  \coordinate (CBIGED)    at ($(CBIGEC)  + ( 0.0              , \bInDistBH     )$) (CBIGED)  [\debugPoint];%
  \coordinate (CBIGF)     at ($(CBIFF)   + ( \bInDistBW       , 0.0            )$) (CBIGF)   [\debugPoint];%
  
  \coordinate (CBIHA)     at ($(CBIGA)   + ( \bInSmallWV      , 0.0            )$) (CBIHA)   [\debugPoint];%
  \coordinate (CBIHB)     at ($(CBIGB)   + ( \bInSmallWV      , 0.0            )$) (CBIHB)   [\debugPoint];%
  \coordinate (CBIHC)     at ($(CBIGC)   + ( \bInSmallWV      , 0.0            )$) (CBIHC)   [\debugPoint];%
  \coordinate (CBIHD)     at ($(CBIGD)   + ( \bInSmallWV      , 0.0            )$) (CBIHD)   [\debugPoint];%
  \coordinate (CBIHE)     at ($(CBIEE)   + ( \bInBigWV        , 0.0            )$) (CBIHE)   [\debugPoint];%
  \coordinate (CBIHEA)    at ($(CBIHE)   + ( 0.0              , \bInSmallBAHV  )$) (CBIHEA)  [\debugPoint];%
  \coordinate (CBIHEB)    at ($(CBIHEA)  + ( 0.0              , \bInDistBH     )$) (CBIHEB)  [\debugPoint];%
  \coordinate (CBIHEC)    at ($(CBIHEB)  + ( 0.0              , \bInSmallBAHV  )$) (CBIHEC)  [\debugPoint];%
  \coordinate (CBIHED)    at ($(CBIHEC)  + ( 0.0              , \bInDistBH     )$) (CBIHED)  [\debugPoint];%
  \coordinate (CBIHF)     at ($(CBIEF)   + ( \bInBigWV        , 0.0            )$) (CBIHF)   [\debugPoint];%
  
  \coordinate (CBIIC)     at ($(CBIHC)   + ( \bInDistBW       , 0.0            )$) (CBIIC)   [\debugPoint];%
  \coordinate (CBIID)     at ($(CBIHD)   + ( \bInDistBW       , 0.0            )$) (CBIID)   [\debugPoint];%
  \coordinate (CBIIE)     at ($(CBIHE)   + ( \bInDistBW       , 0.0            )$) (CBIIE)   [\debugPoint];%
  \coordinate (CBIIF)     at ($(CBIHF)   + ( \bInDistBW       , 0.0            )$) (CBIIF)   [\debugPoint];%
  
  \coordinate (CBIJC)     at ($(CBIIC)   + ( \bInSmallWV      , 0.0            )$) (CBIJC)   [\debugPoint];%
  \coordinate (CBIJD)     at ($(CBIID)   + ( \bInSmallWV      , 0.0            )$) (CBIJD)   [\debugPoint];%
  \coordinate (CBIJE)     at ($(CBIIE)   + ( \bInSmallBAWV    , 0.0            )$) (CBIJE)   [\debugPoint];%
  \coordinate (CBIJF)     at ($(CBIIF)   + ( \bInSmallBAWV    , 0.0            )$) (CBIJF)   [\debugPoint];%
  
  \coordinate (CBIKC)     at ($(CBIJC)   + ( \bInDistBW       , 0.0            )$) (CBIKC)   [\debugPoint];%
  \coordinate (CBIKD)     at ($(CBIJD)   + ( \bInDistBW       , 0.0            )$) (CBIKD)   [\debugPoint];%
  \coordinate (CBIKE)     at ($(CBIJE)   + ( \bInDistBW       , 0.0            )$) (CBIKE)   [\debugPoint];%
  \coordinate (CBIKEA)    at ($(CBIKE)   + ( 0.0              , \bInSmallBBHV  )$) (CBIKEA)  [\debugPoint];%
  \coordinate (CBIKEB)    at ($(CBIKEA)  + ( 0.0              , \bInDistBH     )$) (CBIKEB)  [\debugPoint];%
  \coordinate (CBIKF)     at ($(CBIJF)   + ( \bInDistBW       , 0.0            )$) (CBIKF)   [\debugPoint];%
  
  \coordinate (CBILC)     at ($(CBIKC)   + ( \bInSmallWV      , 0.0            )$) (CBILC)   [\debugPoint];%
  \coordinate (CBILD)     at ($(CBIKD)   + ( \bInSmallWV      , 0.0            )$) (CBILD)   [\debugPoint];%
  \coordinate (CBILE)     at ($(CBIIE)   + ( \bInBigWV        , 0.0            )$) (CBILE)   [\debugPoint];%
  \coordinate (CBILEA)    at ($(CBILE)   + ( 0.0              , \bInSmallBBHV  )$) (CBILEA)  [\debugPoint];%
  \coordinate (CBILEB)    at ($(CBILEA)  + ( 0.0              , \bInDistBH     )$) (CBILEB)  [\debugPoint];%
  \coordinate (CBILF)     at ($(CBIIF)   + ( \bInBigWV        , 0.0            )$) (CBILF)   [\debugPoint];%
  
  \coordinate (CBIMC)     at ($(CBILC)   + ( \bInDistBW       , 0.0            )$) (CBIMC)   [\debugPoint];%
  \coordinate (CBIMD)     at ($(CBILD)   + ( \bInDistBW       , 0.0            )$) (CBIMD)   [\debugPoint];%
  
  \coordinate (CBINC)     at ($(CBIMC)   + ( \bInSmallWV      , 0.0            )$) (CBINC)   [\debugPoint];%
  \coordinate (CBIND)     at ($(CBIMD)   + ( \bInSmallWV      , 0.0            )$) (CBIND)   [\debugPoint];%
  
  \coordinate (CNOAA)     at ($(CBOAA)   + ( \bOutSmallW*0.5 , \bOutSmallH*0.5 )$) (CNOAA)   [\debugPoint];%
  \coordinate (CNOAB)     at ($(CBOAC)   + ( \bOutSmallW*0.5 , \bOutSmallH*0.5 )$) (CNOAB)   [\debugPoint];%
  \coordinate (CNOAC)     at ($(CBOAE)   + ( \bOutSmallW*0.5 , \bOutSmallH*0.5 )$) (CNOAC)   [\debugPoint];%
  \coordinate (CNOBA)     at ($(CBOCA)   + ( \bOutSmallW*0.5 , \bOutSmallH*0.5 )$) (CNOBA)   [\debugPoint];%
  \coordinate (CNOBB)     at ($(CBOCC)   + ( \bOutSmallW*0.5 , \bOutSmallH*0.5 )$) (CNOBB)   [\debugPoint];%
  \coordinate (CNOBC)     at ($(CBOCE)   + ( \bOutSmallW*0.5 , \bOutSmallH*0.5 )$) (CNOBC)   [\debugPoint];%
  
  \coordinate (CNICAAA)   at ($(CBIAA)   + ( \bInSmallWH     , \bInSmallHH     )$) (CNICAAA) [\debugPoint];%
  \coordinate (CNICAAB)   at ($(CBIAC)   + ( \bInSmallWH     , \bInSmallHH     )$) (CNICAAB) [\debugPoint];%
  
  \coordinate (CNICABA)   at ($(CBICA)   + ( \bInSmallWH     , \bInSmallHH     )$) (CNICABA) [\debugPoint];%
  \coordinate (CNICABB)   at ($(CBICC)   + ( \bInSmallWH     , \bInSmallHH     )$) (CNICABB) [\debugPoint];%
  
  \coordinate (CNICACA)   at ($(CBIEA)   + ( \bInSmallWH     , \bInSmallHH     )$) (CNICACA) [\debugPoint];%
  \coordinate (CNICACB)   at ($(CBIEC)   + ( \bInSmallWH     , \bInSmallHH     )$) (CNICACB) [\debugPoint];%
  
  \coordinate (CNICADA)   at ($(CBIGA)   + ( \bInSmallWH     , \bInSmallHH     )$) (CNICADA) [\debugPoint];%
  \coordinate (CNICADB)   at ($(CBIGC)   + ( \bInSmallWH     , \bInSmallHH     )$) (CNICADB) [\debugPoint];%
  
  \coordinate (CNICAEB)   at ($(CBIIC)   + ( \bInSmallWH     , \bInSmallHH     )$) (CNICAEB) [\debugPoint];%
  
  \coordinate (CNICAFB)   at ($(CBIKC)   + ( \bInSmallWH     , \bInSmallHH     )$) (CNICAFB) [\debugPoint];%
  
  \coordinate (CNICAGB)   at ($(CBIMC)   + ( \bInSmallWH     , \bInSmallHH     )$) (CNICAGB) [\debugPoint];%
  
  \coordinate (CNICBA)    at ($(CBIAE)   + ( \bInSmallBAWH   , \bInBigHH       )$) (CNICBA)  [\debugPoint];%
  \coordinate (CNICBBA)   at ($(CBICE)   + ( \bInSmallBBWH   , \bInSmallBBHH   )$) (CNICBBA) [\debugPoint];%
  \coordinate (CNICBBB)   at ($(CBICEB)  + ( \bInSmallBBWH   , \bInSmallBBHH   )$) (CNICBBB) [\debugPoint];%
  \coordinate (CNICBC)    at ($(CBIEE)   + ( \bInSmallBAWH   , \bInBigHH       )$) (CNICBC)  [\debugPoint];%
  \coordinate (CNICBDA)   at ($(CBIGE)   + ( \bInSmallBBWH   , \bInSmallBAHH   )$) (CNICBDA) [\debugPoint];%
  \coordinate (CNICBDB)   at ($(CBIGEB)  + ( \bInSmallBBWH   , \bInSmallBAHH   )$) (CNICBDB) [\debugPoint];%
  \coordinate (CNICBDC)   at ($(CBIGED)  + ( \bInSmallBBWH   , \bInSmallBAHH   )$) (CNICBDC) [\debugPoint];%
  \coordinate (CNICBE)    at ($(CBIIE)   + ( \bInSmallBAWH   , \bInBigHH       )$) (CNICBE)  [\debugPoint];%
  \coordinate (CNICBFA)   at ($(CBIKE)   + ( \bInSmallBBWH   , \bInSmallBBHH   )$) (CNICBFA) [\debugPoint];%
  \coordinate (CNICBFB)   at ($(CBIKEB)  + ( \bInSmallBBWH   , \bInSmallBBHH   )$) (CNICBFB) [\debugPoint];%
  
  \coordinate (CNICCA)    at ($(CBIAG)   + ( \bInBigWH       , \bInBigHH       )$) (CNICCA)  [\debugPoint];%
  \coordinate (CNICCB)    at ($(CBICG)   + ( \bInBigWH       , \bInBigHH       )$) (CNICCB)  [\debugPoint];%
  \coordinate (CNICCC)    at ($(CBIEG)   + ( \bInBigWH       , \bInBigHH       )$) (CNICCC)  [\debugPoint];%

  \node[
    draw = \colorOut, 
    rotate = 90,
    align = center, 
    line width = \lineOutW, 
    inner sep = 0pt, 
    outer sep = 0pt, 
    text width = \bOutSmallH, 
    minimum width = \bOutSmallH, 
    minimum height = \bOutSmallW,
  ] (NNOAA) at (CNOAA) {\textbf{Exchange~rate} \\ \textbf{stabilization} \\ \textbf{techniques}};
  \node[
    draw = \colorOut, 
    rotate = 90,
    align = center, 
    line width = \lineOutW, 
    inner sep = 0pt, 
    outer sep = 0pt, 
    text width = \bOutSmallH, 
    minimum width = \bOutSmallH, 
    minimum height = \bOutSmallW,
  ] (NNOAB) at (CNOAB) {\textbf{Exchange~rate} \\ \textbf{arrangements}};
  \node[
    draw = \colorOut, 
    rotate = 90,
    align = center, 
    line width = \lineOutW, 
    inner sep = 0pt, 
    outer sep = 0pt, 
    text width = \bOutSmallH, 
    minimum width = \bOutSmallH, 
    minimum height = \bOutSmallW,
  ] (NNOAC) at (CNOAC) {\textbf{Monetary regimes}};
  \node[
    draw = \colorOut, 
    rotate = 90,
    align = center, 
    line width = \lineOutW, 
    inner sep = 0pt, 
    outer sep = 0pt, 
    text width = \bOutSmallH, 
    minimum width = \bOutSmallH, 
    minimum height = \bOutSmallW,
  ] (NNOBA) at (CNOBA) {"Which types of \\ practical techniques \\ are used to \\ achieve stability?"};
  \node[
    draw = \colorOut, 
    rotate = 90,
    align = center, 
    line width = \lineOutW, 
    inner sep = 0pt, 
    outer sep = 0pt, 
    text width = \bOutSmallH, 
    minimum width = \bOutSmallH, 
    minimum height = \bOutSmallW,
  ] (NNOBB) at (CNOBB) {"In what way can the \\ value of a crypto- \\ currency be linked \\ to external currencies?"};
  \node[
    draw = \colorOut, 
    rotate = 90,
    align = center, 
    line width = \lineOutW, 
    inner sep = 0pt, 
    outer sep = 0pt, 
    text width = \bOutSmallH, 
    minimum width = \bOutSmallH, 
    minimum height = \bOutSmallW,
  ] (NNOBC) at (CNOBC) {"What is the \\ stabilization target?"}; 
  
  \draw[line width=\lineOutW, \colorOut, -] (CBOEA) rectangle (CBOFB);
  \draw[line width=\lineOutW, \colorOut, -] (CBOEC) rectangle (CBOFD);
  \draw[line width=\lineOutW, \colorOut, -] (CBOEE) rectangle (CBOFF);
  
  \node[
    draw = \colorInB,
    align = center,
    line width = \lineInW,
    inner sep = 0pt,
    outer sep = 0pt,
    text width = \bInSmallWV,
    minimum width = \bInSmallWV,
    minimum height = \bInSmallHV,
  ] (NNICAAA) at (CNICAAA) {\small{}Open \\ market \\ operations \\ (proper)};
  \node[
    draw = \colorInB,
    line width = \lineInW,
    inner sep = 0pt,
    outer sep = 0pt,
    align = center,
    text width = \bInSmallWV,
    minimum width = \bInSmallWV,
    minimum height = \bInSmallHV,
  ] (NNICAAB) at (CNICAAB) {\small{}Direct \\ collaterali- \\ zation};
  \node[
    draw = \colorInB,
    line width = \lineInW,
    inner sep = 0pt,
    outer sep = 0pt,
    align = center,
    text width = \bInSmallWV,
    minimum width = \bInSmallWV,
    minimum height = \bInSmallHV,
  ] (NNICABA) at (CNICABA) {\small{}Open \\ market \\ operations \\ (proxy)};
  \node[
    draw = \colorInB,
    line width = \lineInW,
    inner sep = 0pt,
    outer sep = 0pt,
    align = center,
    text width = \bInSmallWV,
    minimum width = \bInSmallWV,
    minimum height = \bInSmallHV,
  ] (NNICABB) at (CNICABB) {\small{}Proxy \\ collaterali- \\ zation};
  \node[
    draw = \colorInB,
    line width = \lineInW,
    inner sep = 0pt,
    outer sep = 0pt,
    align = center,
    text width = \bInSmallWV,
    minimum width = \bInSmallWV,
    minimum height = \bInSmallHV,
  ] (NNICACA) at (CNICACA) {\small{}Open \\ market \\ operations \\ (self)};
  \node[
    draw = \colorInB,
    line width = \lineInW,
    inner sep = 0pt,
    outer sep = 0pt,
    align = center,
    text width = \bInSmallWV,
    minimum width = \bInSmallWV,
    minimum height = \bInSmallHV,
  ] (NNICACB) at (CNICACB) {\small{}Self \\ collaterali- \\ zation};
  \node[
    draw = \colorInB,
    line width = \lineInW,
    inner sep = 0pt,
    outer sep = 0pt,
    align = center,
    text width = \bInSmallWV,
    minimum width = \bInSmallWV,
    minimum height = \bInSmallHV,
  ] (NNICADA) at (CNICADA) {\small{}Currency \\ interventions};
  \node[
    draw = \colorInB,
    line width = \lineInW,
    inner sep = 0pt,
    outer sep = 0pt,
    align = center,
    text width = \bInSmallWV,
    minimum width = \bInSmallWV,
    minimum height = \bInSmallHV,
  ] (NNICADB) at (CNICADB) {\small{}Interest \\ rates on \\ loans};
  \node[
    draw = \colorInB,
    line width = \lineInW,
    inner sep = 0pt,
    outer sep = 0pt,
    align = center,
    text width = \bInSmallWV,
    minimum width = \bInSmallWV,
    minimum height = \bInSmallHV,
  ] (NNICAEB) at (CNICAEB) {\small{}Interest \\ rates on \\ deposits};
  \node[
    draw = \colorInB,
    line width = \lineInW,
    inner sep = 0pt,
    outer sep = 0pt,
    align = center,
    text width = \bInSmallWV,
    minimum width = \bInSmallWV,
    minimum height = \bInSmallHV,
  ] (NNICAFB) at (CNICAFB) {\small{}Dyn. \\ mining \\ reward};
  \node[
    draw = \colorInB,
    line width = \lineInW,
    inner sep = 0pt,
    outer sep = 0pt,
    align = center,
    text width = \bInSmallWV,
    minimum width = \bInSmallWV,
    minimum height = \bInSmallHV,
  ] (NNICAGB) at (CNICAGB) {\small{}Dyn. burned\\ transaction \\ fee};
  \node[
    draw = \colorInA,
    rotate = 90,
    line width = \lineInAW,
    inner sep = 0pt,
    outer sep = 0pt,
    align = center,
    text width = \bInBigHV,
    minimum width = \bInBigHV,
    minimum height = \bInSmallBAWV,
  ] (NNICBA) at (CNICBA) {Hard peg};
  \node[
    draw = \colorInB,
    line width = \lineInBW,
    inner sep = 0pt,
    outer sep = 0pt,
    align = center,
    text width = \bInSmallBBWV,
    minimum width = \bInSmallBBWV,
    minimum height = \bInSmallBBHV,
  ] (NNICBBA) at (CNICBBA) {No legal tender};
  \node[
    draw = \colorInB,
    line width = \lineInBW,
    inner sep = 0pt,
    outer sep = 0pt,
    align = center,
    text width = \bInSmallBBWV,
    minimum width = \bInSmallBBWV,
    minimum height = \bInSmallBBHV,
  ] (NNICBBB) at (CNICBBB) {Currency board};
  \node[
    draw = \colorInA,
    rotate = 90,
    line width = \lineInAW,
    inner sep = 0pt,
    outer sep = 0pt,
    align = center,
    text width = \bInBigHV,
    minimum width = \bInBigHV,
    minimum height = \bInSmallBAWV,
  ] (NNICBC) at (CNICBC) {Soft peg};
  \node[
    draw = \colorInB,
    line width = \lineInBW,
    inner sep = 0pt,
    outer sep = 0pt,
    align = center,
    text width = \bInSmallBBWV,
    minimum width = \bInSmallBBWV,
    minimum height = \bInSmallBAHV,
  ] (NNICBDA) at (CNICBDA) {Conventional};
  \node[
    draw = \colorInB,
    line width = \lineInBW,
    inner sep = 0pt,
    outer sep = 0pt,
    align = center,
    text width = \bInSmallBBWV,
    minimum width = \bInSmallBBWV,
    minimum height = \bInSmallBAHV,
  ] (NNICBDB) at (CNICBDB) {Horizontal bands};  
  \node[
    draw = \colorInB,
    line width = \lineInBW,
    inner sep = 0pt,
    outer sep = 0pt,
    align = center,
    text width = \bInSmallBBWV,
    minimum width = \bInSmallBBWV,
    minimum height = \bInSmallBAHV,
  ] (NNICBDC) at (CNICBDC) {Crawling};  
  \node[
    draw = \colorInA,
    rotate = 90,
    line width = \lineInAW,
    inner sep = 0pt,
    outer sep = 0pt,
    align = center,
    text width = \bInBigHV,
    minimum width = \bInBigHV,
    minimum height = \bInSmallBAWV,
  ] (NNICBE) at (CNICBE) {Floating \\ arrangements};
  \node[
    draw = \colorInB,
    line width = \lineInBW,
    inner sep = 0pt,
    outer sep = 0pt,
    align = center,
    text width = \bInSmallBBWV,
    minimum width = \bInSmallBBWV,
    minimum height = \bInSmallBBHV,
  ] (NNICBFA) at (CNICBFA) {Free float};
  \node[
    draw = \colorInB,
    line width = \lineInBW,
    inner sep = 0pt,
    outer sep = 0pt,
    align = center,
    text width = \bInSmallBBWV,
    minimum width = \bInSmallBBWV,
    minimum height = \bInSmallBBHV,
  ] (NNICBFB) at (CNICBFB) {Floating with \\ interventions};
  \node[
    draw = \colorInB,
    line width = \lineInW,
    inner sep = 0pt,
    outer sep = 0pt,
    align = center,
    text width = \bInBigWH,
    minimum width = \bInBigWV,
    minimum height = \bInBigHV,
  ] (NNICCA) at (CNICCA) {Exchange rate targeting};
  \node[
    draw = \colorInB,
    line width = \lineInW,
    inner sep = 0pt,
    outer sep = 0pt,
    align = center,
    text width = \bInBigWH,
    minimum width = \bInBigWV,
    minimum height = \bInBigHV,
  ] (NNICCB) at (CNICCB) {Monetary targeting};
  \node[
    draw = \colorInB,
    line width = \lineInW,
    inner sep = 0pt,
    outer sep = 0pt,
    align = center,
    text width = \bInBigWH,
    minimum width = \bInBigWV,
    minimum height = \bInBigHV,
  ] (NNICCC) at (CNICCC) {Inflation targeting};
\end{tikzpicture}
\end{figure*}

\twocolumn
  
\cleardoublepage

\printbibliography

\end{document}